\documentclass[12pt]{article}
\usepackage{a4wide}
\usepackage{axodraw}
\usepackage{latexsym}

\usepackage{pslatex}
\usepackage[latin1]{inputenc}
\usepackage[T1]{fontenc}

\def\bq{\begin{eqnarray}}
\def\eq{\end{eqnarray}}
\def\l{\langle}
\def\r{\rangle} 
\def\eps{\varepsilon}

\def\kp{p_4}
\def\km{p_5}
\def\nf{n_f}
\def\dim{d}
\def\qb{\bar q}
\def\kq{{p_1}}
\def\kqb{{p_3}}
\def\kg{{p_2}}
\def\x#1{{x_{#1}}}
\def\idotp#1,#2{(#1\cdot#2)}
\def\dotp#1{\expandafter\idotp#1}
\def\slash#1{{/\!\!\!\!\!#1}}
\def\osps(#1,#2,#3){\langle#1|\slash{#2}|#3\rangle}
\def\ospsl(#1,#2,#3,#4,#5){\langle#1|\slash{#2}\slash{#3}
\slash{#4}|#5\rangle}
\def\Polg{{\epsilon_g}}

\def\nn{\nonumber}
\def\msbar{{\overline{\rm MS}}}
\def\as{{\alpha_s}}

\begin{document}
\thispagestyle{empty}

\begin{flushright}
  TTP 02-13 \\
  UPRF-2002-09
\end{flushright}

\vspace{1.5cm}

\begin{center}
  {\Large\bf Two-loop amplitudes with nested sums:
    Fermionic contributions to 
             $e^+e^-\to q \qb g$ \\}
  \vspace{1cm}
  {\large Sven Moch$^a$, Peter Uwer$^a$ and Stefan Weinzierl$^b$}\\
  \vspace{1cm}
  $^a${\small {\em Institut f{\"u}r Theoretische Teilchenphysik,
  Universit{\"a}t Karlsruhe}}\\
  {\small {\em 76128 Karlsruhe, Germany}}\\[2mm]
  $^b${\small {\em Dipartimento di Fisica, Universit\`a di Parma,\\
       INFN Gruppo Collegato di Parma, 43100 Parma, Italy}} \\
\end{center}

\vspace{2cm}

% abstract ---------------------------------------
\begin{abstract}\noindent
  {%
    We present the calculation of the 
    $\nf$-contributions to the two-loop
    amplitude for $e^+e^-\to q \qb g$ and 
    give results for the full one-loop amplitude to order $\eps^2$ 
    in the dimensional regularization parameter. 
    Our results agree with those recently obtained by Garland et al.. 
    The calculation makes extensive use of an efficient method based on
    nested sums to calculate two-loop integrals with arbitrary powers 
    of the propagators. 
    The use of nested sums leads in a natural way to multiple polylogarithms 
    with simple arguments, 
    which allow a straightforward analytic continuation.
   }
\end{abstract}

\vspace*{\fill}

% main text ------------------------------------
\newpage

\reversemarginpar

\section{Introduction}
\label{sec:intro}
The last thirty years of experimental studies at colliders together 
with the theoretical investigations, have taught us 
that perturbative quantum chromodynamics (QCD)
gives an excellent description of short-distance scattering 
of strongly interacting partons.
Indeed, today the theory has reached a sufficient maturity that 
it is no longer the target of experimental studies,
but rather a tool in the search for new physics beyond the standard
model.

The search for new physics in particle physics, being 
pursued at present and upcoming collider experiments at the Tevatron and the LHC, 
rely on our ability to make precise predictions for QCD and QCD-associated
processes. The accuracy reached already in present collider
experiments demands next-to-next-to-leading (NNLO) theoretical
predictions within the framework of perturbation theory.
For example the strong coupling constant $\alpha_s$, whose
precise value affects many cross sections, can be measured by using the
data for $e^+ e^- \rightarrow \mbox{3 jets}$.
At present, the error on the extraction of $\alpha_s$ from this measurement
is dominated by theoretical uncertainties \cite{Dissertori:2001mv},
among the main sources there being the 
truncation of the perturbative expansion at a fixed order.
Up to now, event shapes in 3-jet events have been calculated
at next-to-leading order (NLO) for massless
%\cite{Ellis:1981wv,Fabricius:1981sx,Kunszt:1989km,Giele:1992vf,Catani:1997vz}
\cite{Ellis:1981wv}-\cite{Catani:1997vz}
and massive quarks 
%\cite{Rodrigo:1997gy,Rodrigo:1999qg,Bernreuther:1997jn,Brandenburg:1998pu,Nason:1998nw}.
\cite{Rodrigo:1997gy}-\cite{Nason:1998nw}.
To reduce the theoretical uncertainties, it is necesarry
to extend the calculation for massless quarks to
next-to-next-to-leading order.
The calculation of $e^+ e^- \rightarrow \mbox{3 jets}$ at NNLO
requires the tree-level amplitudes for
$e^+ e^- \rightarrow \mbox{5 partons}$ \cite{Berends:1989yn,Hagiwara:1989pp}, 
the one-loop amplitudes for
$e^+ e^- \rightarrow \mbox{4 partons}$ 
%\cite{Bern:1997ka,Bern:1997sc,Glover:1997eh,Campbell:1997tv}
\cite{Bern:1997ka}-\cite{Campbell:1997tv}
as well as the two-loop amplitude
for $e^+e^-\to q \qb g$ together with the one-loop
amplitude $e^+e^-\to q \qb g$ to order $\eps^2$ 
in the dimensional regularization parameter.

While for inclusive quantities like the total hadronic
cross section in $e^+e^-$-annihilation even higher orders have been calculated 
in the past 
%\cite{Gorishnii:1991vf,Surguladze:1991tg,Baikov:2001aa}, 
\cite{Gorishnii:1991vf}-\cite{Baikov:2001aa}, 
the calculation of two-loop four-point scattering amplitudes has been the 
main obstacle for a long time. 
Due to tremendous activity in that field during the past three years 
\cite{Giele:2002hx}, this problem can be considered to be solved --
at least for the case of massless internal quarks and only one external
massive leg. 

The by now more or less standard approach to calculate two-loop
four-point functions has been inspired by the techniques developed
in the calculation of two-point functions. 
The starting point is a reduction
of the tensor integrals through Schwinger parametrization 
%\cite{Tarasov:1996br,Tarasov:1997kx,Anastasiou:1999bn}. 
\cite{Tarasov:1996br}-\cite{Anastasiou:1999bn}. 
This yields immediately scalar integrals with a higher dimension and raised
powers of the propagators. The usual way to proceed then, is to apply
repeatedly algebraic relations between these integrals, 
which follow from Poincare- and Lorentz-invariance 
%\cite{'tHooft:1972fi,Chetyrkin:1981qh,Gehrmann:1999as}. 
\cite{'tHooft:1972fi}-\cite{Gehrmann:1999as}. 
Finally one ends up with a small set of so-called master
integrals which must be solved analytically. 
While for simple topologies it is often straightforward to find the 
reduction scheme (i.e. `triangle rule'), in general it is a non-trivial task to solve
this problem.

In a recent publication \cite{Moch:2001zr,Moch:2001bi}, we have therefore 
proposed a different method to attack this problem. The basic idea is the
following. We solve 
the scalar integrals in higher dimensions and with raised powers of propagators
directly in terms of nested sums instead of reducing all the integrals to 
a small set of master integrals.  
The aim of this paper is to illustrate this method in the calculation 
of the fermionic contributions to the two-loop amplitude $e^+e^-\to q \qb g$, 
i.e. the contributions proportional to the number of quark flavours $\nf$.
We present our results in terms of multiple polylogarithms 
%\cite{Goncharov,Borwein,Remiddi:1999ew}, 
\cite{Goncharov}-\cite{Remiddi:1999ew}, 
which arise naturally from the use of nested sums.
In addition, we show that these multiple polylogarithms can easily be
continued analytically. As a consequence 
the amplitudes for $e^+e^-\to q \qb g$ presented here can also be used 
for $(2+1)$-jet production
in deep-inelastic scattering and the production of a vector boson
($W$, $Z$ or Drell-Yan pair) in hadron-hadron collisions.
The respective amplitudes can be obtained by the crossing symmetry and
simple coupling constant modifications.

The outline of the paper is as follows. In section \ref{sec:prelim} we
present a few properties of the two-loop amplitude. In particular, we
discuss the kinematics, the ultraviolet (UV) renormalization, and the
structure of the soft and collinear singularities. In section 
\ref{sec:method} we outline the calculation. In the following section
\ref{sec:results} we present our results and compare them with those 
recently obtained by Garland et al. \cite{Garland:2002ak}. 
We give our conclusions in section \ref{sec:conclusions}. 
Appendix \ref{sec:appendixresults} contains the results for 
the one-loop amplitude to order $\epsilon^2$ and 
appendix \ref{sec:analyticcont} summarizes properties of 
multiple polylogarithms under analytic continuation.

\section{Preliminaries}
\label{sec:prelim}
\subsection{Kinematics}
In the following we study the reaction
\begin{equation}
  e^+ + e^- \rightarrow  q + g + \bar{q}\, .
\end{equation}
We treat all quarks as massless, that means we work in QCD with $\nf$
massless quark flavours.
To be consistent with earlier work 
\cite{Bern:1997ka,Bern:1997sc} we calculate the amplitude
for the reaction with all particles in the final state
\begin{equation}
  0   \rightarrow  q(p_1) + g(p_2) + \bar{q}(p_3) + e^-(\kp) +  e^+(\km). 
\end{equation}

The kinematical invariants are denoted by
\begin{eqnarray}
  s_{ij} = \left( p_i + p_j \right)^2, \;\;\; 
  s_{ijk} = \left( p_i + p_j + p_k \right)^2,\;\;\;
  s=s_{123}\, ,
\end{eqnarray}
and it is convenient to introduce the dimensionless quantities
\begin{eqnarray}
  x_1 = \frac{s_{12}}{s_{123}}, \;\;\; x_2 = \frac{s_{23}}{s_{123}}.
\end{eqnarray}
For pure photon exchange, the complete amplitude $\cal A_\gamma$ 
for $e^+e^-\to q\qb g$ can be written as the product of a leptonic
current $L_\mu$ with the hadronic current $H_\mu$:
\begin{equation}
  {\cal A_{\gamma}}  
  = -{i\over s} e^2 Q^q T_{\bar i  i}^a  L_\mu H^\mu
  \equiv -e^2 Q^q T_{\bar i  i}^a A_{\gamma},\label{eq:amp-master}
\end{equation}
with $Q^q$ denoting the electric charge of the outgoing quarks in
units of the elementary charge $e=\sqrt{4\pi \alpha}$. The
generator of the $\mbox{SU}(N)$ gauge group is given by $T^a$. The indices
$\bar i, i$ and $a$ describe the color of the outgoing quarks and gluon.
The normalization of the color matrices
is taken to be
\begin{eqnarray}
\mbox{Tr}\left( T^a T^b \right) & = & \frac{1}{2} \delta^{ab}.
\end{eqnarray}
The leptonic current is given by
\begin{equation}
  L_\mu = \bar u(\kp)\gamma_\mu v(\km) ,
\end{equation}
with $u$, $v$ denoting the spinors of the outgoing leptons. 
As we will show later, it is sufficient to consider pure photon
exchange: Working in a helicity basis, the pure photon
exchange amplitude $\cal A_\gamma$ allows the reconstruction 
of the full amplitude including $Z$-boson exchange by adjusting 
the couplings. Using the
anti-commutation relations for the $\gamma$-matrices, one can always achieve
the following form of the hadronic current:
\begin{eqnarray}
\label{eq:hadroniccurrent}
  H_\mu
  &=& c_1 \*{1\over  s}\* \osps(\kq,\kg,\kqb) \* \Polg_\mu
  + c_2 \*{1\over s^2}  \* \osps(\kq,\kg,\kqb) \* \dotp{\Polg,\kq} \* 
  \kqb_\mu \nn\\
  &+&
  c_3 \*{1\over s^2}  \* \osps(\kq,\kg,\kqb) \* \dotp{\Polg,\kqb} \* 
  \kq_\mu 
  +   c_4 \*{1\over s^2}  \* \osps(\kq,\kg,\kqb) \* \dotp{\Polg,\kq} \* 
  \kq_\mu \nn\\
  &+&
  c_5 \*{1\over s^2}  \* \osps(\kq,\kg,\kqb) \* \dotp{\Polg,\kqb} \* 
  \kqb_\mu 
  +   c_6 \*{1\over s} \* \langle\kq|\gamma_\mu|\kqb\rangle \*
  \dotp{\Polg,\kq} 
  \nn\\
  &+&   c_7 \*{1\over s} \* \langle\kq|\gamma_\mu|\kqb\rangle
  \* \dotp{\Polg,\kqb} 
  +   c_8 \*{1\over s} \* \osps(\kq,\Polg,\kqb) \*
  \kq_\mu 
  \nn\\
  &+&   c_9 \*{1\over s} \* \osps(\kq,\Polg,\kqb) \* \kqb_\mu 
  +   c_{10}\*{1\over s} \* 
  \langle\kq|\slash{\Polg}\slash{\kg}\gamma_\mu|\kqb\rangle 
  \nn\\
  &+&   c_{11}\*{1\over s} \* \osps(\kq,\Polg,\kqb) \* \kg_\mu 
  +   
  c_{12}\*{1\over s^2}  \* \osps(\kq,\kg,\kqb) \* \dotp{\Polg,\kq}  \* 
  \kg_\mu \nn\\
  &+&
  c_{13}\*{1\over s^2}  \* \osps(\kq,\kg,\kqb) \* \dotp{\Polg,\kqb} \* 
  \kg_\mu 
\, ,
\end{eqnarray}
where we have used $\langle \kq|$ respective $|\kqb\rangle$ as a short-hand 
notation for the spinors $\bar u(\kq)$ and $v(\kqb)$ of the
outgoing quark and anti-quark. The dimensionless functions $c_i$ depend only on
the ratios $x_i$, the spacetime dimension $\dim =4-2\eps$ and 
the renormalization scale $\mu$,
\begin{equation}
  c_i = c_i\left(x_1,x_2,\eps,{\mu \over s}\right).
\end{equation}
Due to various constraints, for example current conservation,  
\begin{equation}
   (\kq_\mu+\kg_\mu+\kqb_\mu) H^\mu = 0\, ,
\end{equation}
the functions $c_i$ are not all independent 
of each other. It can be easily chown that $c_2, c_4, c_6, c_{12}$ are
sufficient to reconstruct all remaining functions. A similar
conclusion has been drawn in ref. \cite{Garland:2002ak}. 
The relations between 
$c_2, c_4, c_6, c_{12}$ and the remaing functions are:
\begin{eqnarray}
\label{eq:constraint1}
  c_3(\x1,\x2) &=& -  c_2(\x2,\x1),\nn\\
  c_5(\x1,\x2) &=& - c_4(\x2,\x1),\nn\\
  c_{13}(\x1,\x2) &=& - c_{12}(\x2,\x1),
\end{eqnarray}
and
\begin{eqnarray}
\label{eq:constraint2}
  c_1 &=& -{1 \over 2} \*(1-\x2)\*c_3 - {1 \over 2} \*(1-\x1)\*c_5
  +{\x1\over \x2}\*c_6 - {1 \over 2} \*(\x1+\x2)\*c_{13},\nn\\
  c_7 &=& -{\x1\over \x2}\*c_6,\nn\\
  c_8 &=& -{1 \over 2}\*\x2 \*c_3-{1 \over 2}\*\x1 \*c_4,\nn\\
  c_9 &=& -{1 \over 2}\*\x1 \*c_2-{1 \over 2} \*\x2 \*c_5,\nn\\
  c_{10} &=& 
  +{1 \over 4} \*(1-\x1)\*(c_2-c_5)
  - {1 \over 4} \*(1-\x2)\*(c_3-c_4)
  +{1 \over 2} \*{(\x1+\x2)\over \x2}\*c_6\nn\\
  & & +{1 \over 4} \*(\x1+\x2)\*(c_{12} - c_{13})
  ,\nn\\
  c_{11} &=& {1 \over 2} \*(1-\x2)\*c_3 + {1 \over 2} \*(1-\x1)\*c_5
  -{\x1\over \x2}\*c_6
  +{1 \over 2} \*\x1\*c_{13} -{1 \over 2}\*\x1 \*c_{12}.
\end{eqnarray}
In the actual calculation we have not used these constraints. 
Instead we calculated all $c_1$--$c_{13}$ and used 
eqs. (\ref{eq:constraint1}) and (\ref{eq:constraint2}) 
as a cross-check on our calculation.

Beyond the leading-order, one encounters UV as well as soft and collinear
singularities. We use dimensional regularization
\cite{'tHooft:1972fi,Bollini:1972ui} to regulate both
types of singularities. There are several variants of dimensional 
regularization which are used in loop calculations in QCD:
Conventional dimensional regularization (CDR) \cite{Collins} continues 
all momenta and all polarization vectors to $\dim $ dimensions.
The 't Hooft-Veltman (HV) scheme \cite{'tHooft:1972fi} takes
the momenta and the helicities of the unobserved particles in 
$\dim $ dimensions,
whereas the momenta and the helicities of the observed particles are  
four-dimensional.
The CDR scheme is often employed within the interference method, but is not
suited for the calculation of amplitudes.
Enforcing the CDR scheme in the calculation of amplitudes requires 
the introduction
of external states with ``$\eps$''-helicities \cite{Kosower:1991ax}.
Furthermore, there are several versions of four-dimensional schemes on the market.
Despite the name ``four-dimensional schemes'', they are variants of dimensional regularization.
The name refers to how these schemes treat unobserved internal particles and the Dirac
algebra.
The four-dimensional helicity scheme (FDH) \cite{Bern:1992aq}
introduces an additional parameter $\dim _s$ for unobserved internal states, 
which is set to $4$ at the end of the calculation.
It has the advantage that it respects supersymmetric Ward identities
up to two loops.
The four-dimensional scheme defined in \cite{Weinzierl:1999xb} keeps
the  Dirac algebra in four
dimensions and allows the use of four-dimensional Fierz- and Schouten 
identities, at the expense
of having to restore Ward identities.
These schemes can lead to considerable simplifications, in particular
if many external particles are involved.
For the process $e^+ e^- \rightarrow 3\;\mbox{jets}$, the number of
external particles is relatively small
and we do not consider these schemes further.

In this paper we keep our calculation rather general and
we decompose the amplitude into spinor strings, which we can compute
without any reference to the dimensionality of the external
polarization vectors. 
Internal particles and the Dirac algebra are treated in $\dim$ dimensions.
From these spinor strings we can easily deduce the results in the CDR scheme
and the HV scheme.
The result in the CDR scheme
is obtained by interfering 
the two-loop amplitude with the Born amplitude.
On the other hand, by contracting with explicit representations of 
polarization vectors, we obtain
helicity amplitudes in the 't Hooft-Veltman scheme. 

Including the
$Z$-boson exchange, the situation becomes slightly more complicated. Due to the
presence of $\gamma_5$, a specific scheme has to be chosen. 
Schemes which allow for a consistent treatment of $\gamma_5$ are for example
the HV scheme or the scheme defined in \cite{Weinzierl:1999xb}.
In both schemes, the regularization procedure violates certain Ward identities,
which have to be restored by finite renormalizations.
However, since the amplitudes considered in this paper do not contain closed
fermion loops with axial-vector couplings, the results for $Z$-boson exchange
can be obtained from the ones for pure photon exchange by a simple adjustment
of the electro-weak couplings.

It should be noted that in general the finite part of the amplitudes are also
scheme-dependent.
As long as soft and collinear singularities are considered, this is not
really an issue. Using the same scheme in the calculation of the
divergent contributions from the real corrections any scheme
dependence cancels out at the end. 
For the UV divergences however, one has to keep in mind that in general the 
coupling constant is scheme dependent. 
The values of the coupling constant in two different schemes are related by a finite
renormalization.
The coupling constants in the CDR scheme and the HV scheme are identical, 
since the internal states in these two schemes are treated in the same way.
The CDR or HV scheme, together with the modified minimal subtraction prescription ($\msbar$),
defines the usual coupling $\alpha^\msbar_s$ and it is most useful to quote our results
in these schemes.

To obtain the results in the 't Hooft-Veltman scheme we work in the helicity basis.
For fermions, spinors of definite helicity are given by:
\begin{eqnarray}
u_{\pm}(p) & = & \frac{1}{2} \left( 1 \pm \gamma_5 \right) u(p), \nonumber \\
v_{\pm}(p) & = & \frac{1}{2} \left( 1 \mp \gamma_5 \right) v(p).
\end{eqnarray}
We introduce the following short-hand notation for spinors with
definite helicity:
\begin{eqnarray}
|i \pm \rangle \,\, =&\!  | p_i \pm \rangle \!&=\,\, u_\pm(p_i) \,=\, v_\mp(p_i), \nonumber \\
\langle i \pm |\,\,  =&\! \langle p_i \pm | \!&=\,\, \bar{u}_\pm(p_i) \,=\, \bar{v}_\mp(p_i),
\end{eqnarray}
and the spinor products are then defined as
\begin{eqnarray}
\langle p q \rangle & = & \langle p - | q + \rangle, \nonumber \\
\left[ p q \right] &= & \langle p + | q - \rangle .
\end{eqnarray}
Gluon polarization vectors are expressed in terms of Weyl spinors as
%\cite{Berends:1981rb,DeCausmaecker:1982bg,Gastmans:1990xh,Gunion:1985vc,Kleiss:1985yh,Xu:1987xb}
\cite{Berends:1981rb}-\cite{Xu:1987xb}
\begin{eqnarray}
\eps^+_\mu(k,q) = \frac{\l q- | \gamma_\mu | k- \r }{\sqrt{2} \l qk \r },
& & 
\eps^-_\mu(k,q) = \frac{\l q+ | \gamma_\mu | k+ \r }{\sqrt{2} [ kq ] },
\end{eqnarray}
where $k$ is the gluon momentum and $q$ is an arbitrary null reference momentum.
The dependence on the reference momentum drops out in final gauge-invariant amplitudes.
An appropriate choice of $q$ can lead to a significant reduction in the number of 
diagrams which need to be evaluated.
\begin{figure}
\begin{center}
\begin{picture}(100,100)(0,0)
 \ArrowLine(0,70)(20,50)
 \ArrowLine(20,50)(0,30)
 \Vertex(20,50){2}
 \Photon(20,50)(70,50){2}{4}
 \ArrowLine(70,50)(100,80)
 \ArrowLine(100,20)(70,50)
 \Gluon(70,50)(100,50){2}{4}
 \GCirc(70,50){10}{0.7}
 \Text(105,80)[l]{$p_1$}
 \Text(105,50)[l]{$p_2$}
 \Text(105,20)[l]{$p_3$}
 \Text(-5,30)[r]{$\kp$}
 \Text(-5,70)[r]{$\km$}
\end{picture}
\caption{\label{diagram3} Labelling of the external legs for the amplitude. 
The lepton pair is denoted by $\kp$ and $\km$.}
\end{center}
\end{figure}
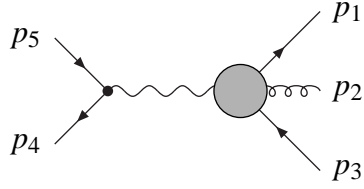

Helicity conservation for massless quarks and leptons ensures that there are
only $2^3=8$ possible helicity configurations for $e^+e^-\to q \qb g$,
namely two choices for each fermion line together with two possible gluon polarizations.
Because the electron line couples through the current
$\l 4 \pm | \gamma^\mu | 5 \pm \r = \l 5 \mp | \gamma^\mu | 4 \mp \r$,
it is trivial to reverse its helicity simply by exchanging 
$\kp \leftrightarrow \km$
and by adjusting the weak couplings.
Parity and charge conjugation can be used to further reduce the number of helicity
amplitudes which need to be calculated.
Parity reverses all helicities simultaneously and is implemented by complex
conjugating all spinor products (e.g. $\l ij \r \leftrightarrow [ ji ]$).
Charge conjugation reverses the arrows of each fermion line. 
In addition there is a factor $(-1)$ for each external gauge boson.
Thus we are left with just one independent helicity amplitude, which we take to
be $A_\gamma(1^+, 2^+, 3^-, 4^+, 5^-)$.

Keeping the reference momentum $q$ in the gluon polarization vector
arbitrary we obtain:
\begin{eqnarray}
  A_\gamma(1^+, 2^+, 3^-, 4^+, 5^-) & = & 
  \frac{i}{\sqrt{2}} \frac{1}{\l q2 \r} \frac{1}{s^2} \left\{
  \right. 
  2 c_1 \l q5 \r [ 42 ] [ 12 ] \l 23 \r
  + c_2 {1\over s} \l q1 \r [ 12 ] [ 43 ] \l 35 \r [ 12 ] \l 23 \r
  \nonumber \\ & &
  + c_3 {1\over s}\l q3 \r [ 32 ] [ 41 ] \l 15 \r [ 12 ] \l 23 \r
  + c_4 {1\over s}\l q1 \r [ 12 ] [ 41 ] \l 15 \r [ 12 ] \l 23 \r
  \nonumber \\ & &
  + c_5 {1\over s}\l q3 \r [ 32 ] [ 43 ] \l 35 \r [ 12 ] \l 23 \r
  + 2 c_6 \l q1 \r [ 12 ] [ 41 ] \l 35 \r 
  \nonumber \\ & &
  + 2 c_7 \l q3 \r [ 32 ] [ 41 ] \l 35 \r 
  + 2 c_8 \l q3 \r [ 12 ] [ 41 ] \l 15 \r 
  \nonumber \\ & &
  + 2 c_9 \l q3 \r [ 12 ] [ 43 ] \l 35 \r 
  + 4 c_{10} \l q2 \r [ 12 ] [ 42 ] \l 35 \r 
  \nonumber \\ & &
  + 2 c_{11} \l q3 \r [ 12 ] [ 42 ] \l 25 \r 
  + c_{12} {1\over s}\l q1 \r [ 12 ] [ 42 ] \l 25 \r [ 12 ] \l 23 \r 
  \nonumber \\ & & \left.
  + c_{13} {1\over s}\l q3 \r [ 32 ] [ 42 ] \l 25 \r [ 12 ] \l 23 \r 
 \right\} .
\end{eqnarray}
Using $q=p_3$ and the constraints of eq. (\ref{eq:constraint2}) 
this can be simplified to:
\begin{eqnarray}
\lefteqn{
  A_\gamma(1^+, 2^+, 3^-, 4^+, 5^-) \, =} 
\nonumber \\ & & 
  \frac{i}{\sqrt{2}} \frac{ [12]}{s^3} \left\{
    s \l 35 \r [42] \left[ (1-x_1) \left( c_2 + \frac{2}{x_2} c_6 - c_{12}
  \right) + (1-x_2) \left( c_4 - c_{12} \right) + 2  c_{12}
 \right]
  \right. \nonumber \\
  & & \left.
    - \l 31 \r [ 12 ] \left[ 
      [ 43 ] \l 35 \r \left( c_2 + \frac{2}{x_2} c_6 - c_{12} \right)
      + [ 41 ] \l 15 \r \left( c_4 - c_{12} \right)
    \right]
  \right\}\, .
\end{eqnarray}
We see, that for this choice of helicities and $q$ the amplitude $A_\gamma$ 
is described by three linear combinations of $c_2, c_4, c_6, c_{12}$.

The perturbative expansion of the functions $c_i$ and $A_\gamma$ is defined through
\begin{eqnarray}
  c_i &=& \sqrt{4 \pi \alpha_s} \left(
    c_i^{(0)} 
    +\left( \frac{\alpha_s}{2\pi}\right) c_i^{(1)} 
    +\left( \frac{\alpha_s}{2\pi}\right)^2 c_i^{(2)}
    + O(\as^3)
  \right), 
\nonumber \\
   A_\gamma &=& \sqrt{4 \pi \alpha_s} \left(
    A_\gamma^{(0)} 
    +\left( \frac{\alpha_s}{2\pi}\right) A_\gamma^{(1)} 
    +\left( \frac{\alpha_s}{2\pi}\right)^2 A_\gamma^{(2)}
    + O(\as^3)
  \right).
\end{eqnarray}
Inserting the leading-order result for the $c_i$ 
\begin{eqnarray}
  c_2^{(0)}=c_4^{(0)}=c_{12}^{(0)}=0,\quad c_6^{(0)} = {2 \over x_1}\, ,
\end{eqnarray}
one gets the following result for the tree amplitude
\begin{eqnarray}
    A^{(0)}_\gamma(1^+, 2^+, 3^-, 4^+, 5^-) & = &
 2 \sqrt{2} i \frac{\l 35 \r^2}{\l 12 \r \l 23 \r \l 45 \r}.
\end{eqnarray}
In the helicity basis one can easily account for the $Z$-boson exchange by
adjusting the couplings:%
\def\hel#1{{\lambda_#1}}
\begin{equation}
  {\cal A}_{\gamma
  Z}(1^{\hel1},2^{\hel2},3^{\hel3},4^{\hel4},5^{\hel5})
  = e^2 (- Q^q + v_{\hel4}^e v_{\hel1}^q {\cal P}_Z(s)) 
  A_\gamma(1^{\hel1},2^{\hel2},3^{\hel3},4^{\hel4},5^{\hel5}) \, ,
\end{equation}
with
\begin{eqnarray}
  {\cal P}_Z(s) = \frac{s}{s-m_Z^2 + i m_Z \Gamma_Z }.
\end{eqnarray}
Here, $m_Z$ and $\Gamma_Z$ are the mass and the width of the $Z$-boson.
The left- and right handed couplings of fermions to the $Z$-boson are
\begin{eqnarray}
  v_-^f = \frac{I_3^f - Q_f \sin^2 \theta_W}{\sin \theta_W \cos \theta_W}, & &
  v_+^f = \frac{- Q_f \sin \theta_W}{\cos \theta_W},
\end{eqnarray}
where $Q_f$ and $I_3^f$ are the charge and the third component of the weak 
isospin of the fermion $f$ and $\theta_W$ is the Weinberg angle.

\subsection{Ultraviolet renormalization}
\label{sec:UV_renorm}
\def\bare{\mbox{\scriptsize \rm bare}}
\def\ren{\mbox{\scriptsize \rm ren}}
\def\fin{\mbox{\scriptsize \rm fin}}

The amplitudes we present are the renormalized ones, i.e. 
the ultraviolet subtraction has been performed.
To obtain the renormalized amplitudes in the $\overline{\mbox{MS}}$ scheme, 
one replaces the bare coupling $\alpha_0$ with the renormalized coupling $\alpha_s(\mu^2)$ evaluated
at the renormalization scale $\mu^2$:
\begin{eqnarray}
\alpha_0 & = & \alpha_s S_\eps^{-1} \left[ 1 
 -\frac{\beta_0}{\eps} \left( \frac{\alpha_s}{2\pi} \right)
 + \left( \frac{\beta_0^2}{\eps^2} - \frac{\beta_1}{2\eps} \right) 
   \left( \frac{\alpha_s}{2\pi} \right)^2
 + {\cal O}(\alpha_s^3) \right],
\end{eqnarray}
where 
\begin{eqnarray}
S_\eps & = & \left( 4 \pi \right)^\eps e^{-\eps\gamma_E} \, ,
\end{eqnarray}
is the typical phase-space volume factor in $\dim =4-2\eps$ dimensions, 
$\gamma_E$ is Euler's constant,
and $\beta_0$ and $\beta_1$ are the first two coeffcients of the QCD $\beta$-function:
\begin{eqnarray}
\beta_0 = \frac{11}{6} C_A - \frac{2}{3} T_R \nf,
&\;\;\;&
\beta_1 = \frac{17}{6} C_A^2 - \frac{5}{3} C_A T_R \nf - C_F T_R \nf,
\end{eqnarray}
with the color factors
\begin{eqnarray}
C_A = N, \;\;\; C_F = \frac{N^2-1}{2N}, \;\;\; T_R = \frac{1}{2}.
\end{eqnarray}

It is convenient here and for the subsequent discussion of the soft and
collinear singularities to introduce a different notation \cite{Catani:1998bh}.
In an orthogonal basis of unit vectors ${|\bar i,i,a\r}$ in the three parton
color space we define an abstract vector $|\cal M\r $  through
\begin{equation}
  {\cal A}  \equiv \l \bar i ,i,a|\cal M\r ,
\end{equation}
with the expansion in the coupling $\as$ defined by
\begin{eqnarray}
  |\cal M\r & = & 
 4\pi \alpha \sqrt{4 \pi \alpha_s} 
 \left[
 \left| {\cal M}^{(0)} \right\r
 + \left( \frac{\alpha_s}{2\pi} \right) \left| {\cal M}^{(1)} \right\r
 + \left( \frac{\alpha_s}{2\pi} \right)^2 \left| {\cal M}^{(2)} \right\r
 + {\cal O}(\alpha_s^3)
 \right].
\end{eqnarray}

Then, the renormalized two-loop amplitude can be expessed as
\begin{eqnarray}
\left| {\cal M}_{\ren} \right\r & = & 
 4\pi \alpha \sqrt{4 \pi \alpha_s} S_\eps^{-1/2}  
 \left[
 \left| {\cal M}^{(0)}_{\ren} \right\r
 + \left( \frac{\alpha_s}{2\pi} \right) \left| {\cal M}^{(1)}_{\ren} \right\r
 + \left( \frac{\alpha_s}{2\pi} \right)^2 \left| {\cal M}^{(2)}_{\ren} \right\r
 + {\cal O}(\alpha_s^3)
 \right], 
\end{eqnarray}
At two loops the relation between the renormalized and the bare amplitudes is
given by
\begin{eqnarray}
\left| {\cal M}^{(0)}_{\ren} \right\r & = & 
 \left| {\cal M}^{(0)}_{\bare} \right\r , \nonumber \\
\left| {\cal M}^{(1)}_{\ren} \right\r & = & 
 S_\eps^{-1} \left| {\cal M}^{(1)}_{\bare} \right\r 
 - \frac{\beta_0}{2\eps} \left| {\cal M}^{(0)}_{\bare} \right\r , \nonumber \\
\left| {\cal M}^{(2)}_{\ren} \right\r & = & 
 S_\eps^{-2} \left| {\cal M}^{(2)}_{\bare} \right\r 
 - \frac{3\beta_0}{2\eps} S_\eps^{-1} \left| {\cal M}^{(1)}_{\bare} \right\r
 + \left(  \frac{3\beta_0^2}{8\eps^2} - \frac{\beta_1}{4\eps} \right) 
  \left| {\cal M}^{(0)}_{\bare} \right\r .
\end{eqnarray}
Thus, we obtain for the renormalized functions $c_{i,\ren}$ 
\begin{eqnarray}
  c_{i}^{(1),\ren} &=& S_\eps^{-1} c_{i}^{(1),\bare}\, ,\nn\\
  c_{i}^{(2),\ren} &=& S_\eps^{-2} c_{i}^{(2),\bare}
  - \frac{3\beta_0}{2\eps} S_\eps^{-1} c_{i}^{(1),\bare}\, ,
\end{eqnarray}
for $i=\{2,4,12\}$ and
\begin{eqnarray}
  c_{6}^{(1),\ren} &=& S_\eps^{-1} c_{6}^{(1),\bare}
  - \frac{\beta_0}{2\eps}c_{6}^{(0)}\, ,\nn\\
  c_{6}^{(2),\ren} &=& S_\eps^{-2} c_{6}^{(2),\bare}
  - \frac{3\beta_0}{2\eps} S_\eps^{-1} c_{6}^{(1),\bare}
  + \left(  \frac{3\beta_0^2}{8\eps^2} - \frac{\beta_1}{4\eps} \right) 
  c_{6}^{(2),\bare}\, .
\end{eqnarray}
In this paper we set the renormalization scale $\mu^2=s$. 
The complete scale dependence is easily recovered by expanding
the prefactor
\bq
\left( \frac{\mu^2}{s} \right)^{l \eps},
\eq
accompanying an $l$-loop amplitude to the appropriate order in $\eps$.

\subsection{Infrared structure}
\label{sec:infrared}

Based on universal properties of soft and collinear limits, 
the infrared pole structure of two-loop amplitudes has been 
predicted by Catani \cite{Catani:1998bh}. 
Here, we briefly review how to organize these infrared poles 
for $e^+e^-\to q \qb g$.
We start with the one-loop amplitude, which can be written as
\begin{eqnarray}
\left| {\cal M}^{(1)} \right\r & = & 
 {\bf I}^{(1)}(\eps) \left| {\cal M}^{(0)} \right\r
 + \left| {\cal F}^{(1)} \right\r .
\end{eqnarray}
Here ${\bf I}^{(1)}(\eps)$ contains all infrared double and single poles in $1/\eps$ and
$|{\cal F}^{(1)}\r$ is a finite remainder.
At two-loops, the corresponding formula reads:
\begin{eqnarray}
\left| {\cal M}^{(2)} \right\r & = & 
 {\bf I}^{(1)}(\eps) \left| {\cal M}^{(1)} \right\r
 + {\bf I}^{(2)}(\eps) \left| {\cal M}^{(0)} \right\r
 + \left| {\cal F}^{(2)} \right\r .
\end{eqnarray}
The one-loop insertion operator ${\bf I}^{(1)}$ is given by
\begin{eqnarray}
{\bf I}^{(1)}(\eps) & = & 
 \frac{1}{2} \frac{1}{\Gamma(1-\eps)} e^{\eps \gamma_E} 
 \sum\limits_{i} \frac{1}{ {\bf T}_i^2} {\cal V}_i(\eps)
 \sum\limits_{j \neq i} {\bf T}_i {\bf T}_j
 \left( \frac{\mu^2}{-s_{ij}} \right)^\eps,
\end{eqnarray}
where 
\begin{eqnarray}
 {\cal V}_i(\eps) & = &
  {\bf T}_i^2 \frac{1}{\eps^2} + \gamma_i \frac{1}{\eps} \, ,
\end{eqnarray}
and the coefficients ${\bf T}_i^2$ and $\gamma_i$ are
\begin{eqnarray}
{\bf T}_q^2 = {\bf T}_{\bar{q}}^2 = C_F,
 & &
{\bf T}_g^2 = C_A, \nonumber \\
\gamma_q = \gamma_{\bar{q}} = \frac{3}{2} C_F,
 & & 
\gamma_g = \beta_0.
\end{eqnarray}
In general, the color operators ${\bf T}_i {\bf T}_j$ give rise to color
correlations. 
However, the color structure for the amplitude  $e^+e^- \to q \qb g$ 
is rather trivial and the color operators are proportional to the identity matrix
in color space:
\begin{eqnarray}
& & {\bf T}_q {\bf T}_{\bar{q}} = T_R {1\over N}, \nonumber \\
& & {\bf T}_q {\bf T}_{g} = {\bf T}_g {\bf T}_{\bar{q}} = - T_R N.
\end{eqnarray}
Explicitly, the one-loop insertion operator reads for $e^+e^- \to q \qb g$:
\begin{eqnarray}
\lefteqn{
{\bf I}^{(1)}(\eps)  =  
 \frac{1}{2} \frac{1}{\Gamma(1-\eps)} e^{\eps \gamma_E} 
 \left( \frac{\mu^2}{-s} \right)^\eps
 T_R } 
 \nonumber \\
 & & \times
 \left[ 
   {2\over N} \left( 1 - x_1 - x_2 \right)^{-\eps}
     \left( \frac{1}{\eps^2} + \frac{3}{2} \frac{1}{\eps} \right) 
   - N \left( x_1^{-\eps} + x_2^{-\eps} \right)
     \left( \frac{2}{\eps^2} + \left( \frac{\beta_0}{C_A} + \frac{3}{2} \right) \frac{1}{\eps} \right)
   \right]\, .
\end{eqnarray}
The two-loop insertion operator has the form 
\begin{eqnarray}
\lefteqn{
{\bf I}^{(2)}(\eps) \, =}
\nonumber \\ &  & 
 - \frac{1}{2} {\bf I}^{(1)}(\eps) \left( {\bf I}^{(1)}(\eps) + 2 \beta_0 \frac{1}{\eps} \right)
 + e^{-\eps \gamma_E} \frac{\Gamma(1-2\eps)}{\Gamma(1-\eps)}
   \left( \beta_0 \frac{1}{\eps} + K \right) {\bf I}^{(1)}(2 \eps)
 + {\bf H}^{(2)},
\end{eqnarray}
where
\begin{eqnarray}
K & = & \left( \frac{67}{18} - \frac{\pi^2}{6} \right) C_A - \frac{10}{9} T_R \nf.
\end{eqnarray}
The function ${\bf H}^{(2)}$ is process- and scheme-dependent and for 
$e^+e^- \to q \qb g$, it is given by \cite{Garland:2001tf,Bern:2002tk}
\begin{eqnarray}
 {\bf H}^{(2)} & = & \frac{1}{4} \frac{1}{\Gamma(1-\eps)} e^{\eps \gamma_E}
   \frac{1}{\eps}
   \left( {\bf H}_q^{(2)} + {\bf H}_g^{(2)} + {\bf H}_{\bar{q}}^{(2)} \right),
\end{eqnarray}
where ${\bf H}_q^{(2)} = {\bf H}_{\bar{q}}^{(2)}$ and
\begin{eqnarray}
 \nonumber \\
{\bf H}_q^{(2)} & = & 
\left( \frac{7}{4} \zeta_3 - \frac{11}{96} \pi^2 + \frac{409}{864} \right) N^2
+ \left( - \frac{1}{4} \zeta_3 - \frac{\pi^2 }{96} - \frac{41}{108} \right)
+ \left( - \frac{3}{2} \zeta_3 + \frac{\pi^2 }{8} - \frac{3}{32} \right) 
   \frac{1}{ N^2}
\nonumber \\
 & & 
 + \left( \frac{\pi^2}{48} - \frac{25}{216} \right) \frac{N^2 -1}{N} \nf,
 \nonumber \\
{\bf H}_g^{(2)} & = & 
\left( \frac{\zeta_3}{2} + \frac{11}{144} \pi^2 + \frac{5}{12} \right) N^2
+ \left( - \frac{\pi^2}{72} - \frac{89}{108} \right) N \nf
- \frac{\nf}{4N}
+ \frac{5}{27} \, \nf^2. 
\end{eqnarray}
Using the above results we define the finite functions
$c_{i}^{{(j),\fin}}$:
\begin{eqnarray}
  c_{i}^{(1),\fin} &=& c_{i}^{(1),\ren},\nn\\
  c_{i}^{(2),\fin} &=& c_{i}^{(1),\ren} - {\bf I}^{(1)}(\eps) c_{i}^{(1),\ren}
  \label{eq:def-results1}
\end{eqnarray}
for $i=\{2,4,12\}$ and
\begin{eqnarray}
   c_{6}^{(1),\fin} &=& c_{6}^{(1),\ren} - {\bf I}^{(1)}(\eps) c_{6}^{(0)},\nn\\
   c_{6}^{(2),\fin} &=& c_{6}^{(1),\ren} - {\bf I}^{(1)}(\eps) c_{6}^{(1),\ren}
   - {\bf I}^{(2)}(\eps) c_{6}^{(0)}.
  \label{eq:def-results2}
\end{eqnarray}
Explicit results for the functions $c_{i}^{(2),\fin}$ 
are given in section \ref{sec:results} and for $c_{i}^{(1),\fin}$ in the 
appendix \ref{sec:appendixresults}.

\section{Method of calculation}
\label{sec:method}

In this section, we will discuss the method to calculate the virtual
amplitudes for $e^+e^-\to q \qb g$.
We have used QGRAF \cite{Nogueira:1991ex} for the generation of all 
Feynman diagrams, which contribute to the process $e^+e^-\to q \qb g$ 
up to two loops.

The evaluation of the diagrams 
leads to tensor integrals, which multiply the various spinor structures of
eq. (\ref{eq:hadroniccurrent}).
Note, that there is no need here to consider projectors for the various spinor
coefficients.

The tensor integrals are mapped to combinations of
scalar integrals with higher powers of propagators and different values of $\dim $ 
%\cite{Tarasov:1996br,Tarasov:1997kx,Anastasiou:1999bn}. 
\cite{Tarasov:1996br}-\cite{Anastasiou:1999bn}. 
For this purpose, one introduces Schwinger parameters for the propagators 
\begin{eqnarray}
\frac{1}{(-k^2)^\nu} & = & \frac{1}{\Gamma(\nu)} \int\limits_0^\infty dx \;
x^{\nu-1} \exp(x k^2).
\end{eqnarray}
Combining the exponentials arising from different propagators one
obtains a quadratic form in the loop momenta. 
For instance, for a given two-loop integral with loop momenta $k_1$ and $k_2$, 
one has then 
\begin{eqnarray}
I(d,\nu_1,\ldots,\nu_k)&=&
\int \frac{d^{\dim }k_1}{i\pi^{\dim /2}}
\int \frac{d^{\dim }k_2}{i\pi^{\dim /2}}
\frac{1}{(-k_1^2)^{\nu_1}\ldots (-k_n^2)^{\nu_n}}\nn\\
&=&
\int \frac{d^{\dim }k_1}{i\pi^{\dim /2}}
\int \frac{d^{\dim }k_2}{i\pi^{\dim /2}} \left( \prod_{i=1}^n \frac{1}{\Gamma(\nu_i)}
\int^{\infty}_0 dx_i\, x_i^{\nu_i-1} \right) \exp\left ( \sum_{i=1}^n x_i k_i^2 \right),
\end{eqnarray}
and
\begin{equation}
\label{eq:quadraticform}
\sum_{i=1}^n x_i k_i^2 = a\, k_1^2 + b\, k_2^2 + 2\, c \,k_1 \cdot k_2 + 2\, d
\cdot k_1 + 2 \, e \cdot k_2 + f\ .
\end{equation}
The momenta $k_3,\ldots,k_k$ are 
linear combinations of the loop momenta $k_1,k_2$ and the external momenta. 
The coefficients $a$, $b$, $c$, $d^\mu$, $e^\mu$ and $f$ are directly 
readable from the actual graph: $a (b) = \sum x_i$, where the sum 
runs over the legs in the $k_1$ $(k_2)$ loop, and $c = \sum x_i$ with 
the sum running over
the legs common to both loops. 
With a suitable change of variables for the loop momenta $k_1, k_2$, one can 
diagonalize the quadratic form 
and the momentum integration can be performed as Gaussian integrals 
over the shifted loop momenta according to
\begin{eqnarray}
\int \frac{d^{\dim }k}{i\pi^{\dim /2}} \exp \left( {\cal P} k^2 \right) & = & 
 \frac{1}{{\cal P}^{\dim /2}}.
\end{eqnarray}

Lorentz invariance allows immediately to relate the following symmetric
tensor integrals to scalar integrals:
\begin{eqnarray}
\int \frac{d^{\dim }k}{i\pi^{\dim /2}} k^\mu k^\nu f(k^2) & = & 
 \frac{1}{\dim } g^{\mu\nu} \int \frac{d^{\dim }k}{i\pi^{\dim /2}} k^2 f(k^2), \nonumber \\
\int \frac{d^{\dim }k}{i\pi^{\dim /2}} k^\mu k^\nu k^\rho k^\sigma f(k^2) & = & 
 \frac{1}{\dim (\dim +2)} 
  \left( g^{\mu\nu} g^{\rho\sigma} + g^{\mu\rho} g^{\nu\sigma} + g^{\mu\sigma} g^{\nu\rho} \right) 
  \int \frac{d^{\dim }k}{i\pi^{\dim /2}} k^2 f(k^2), 
\end{eqnarray}
and the generalization to arbitrary higher tensor structures is obvious.

In the remaining Schwinger parameter integrals, the tensor integrals 
introduce additional factors of the parameters $x_i$ and of $1/{\cal P}$.
These additional factors can be absorbed into scalar integrals with 
higher powers of propagators and shifted dimensions, by introducing 
operators ${\bf i}^+$,  which raise the power of propagator
$i$ by one, or an operator ${\bf d}^+$ that increases the dimension by two,
\begin{eqnarray}
\nu_i {\bf i}^+ \frac{1}{\left(-k_i^2\right)^{\nu_i}} & = & 
 \nu_i \frac{1}{\left(-k_i^2\right)^{\nu_i+1}}
 = \frac{1}{\Gamma(\nu_i)} \int\limits_0^\infty dx_i \; x_i^{\nu_i-1} x_i \exp(x_i k^2),
 \nonumber \\
{\bf d}^+ \int \frac{d^{\dim }k}{i\pi^{\dim /2}} \exp \left( {\cal P} k^2 \right) & = &
 \int \frac{d^{(\dim +2)}k}{i\pi^{(\dim +2)/2}} \exp \left( {\cal P} k^2 \right)
 = \frac{1}{{\cal P}} \frac{1}{{\cal P}^{\dim /2}}.
\end{eqnarray}

\begin{figure}
\begin{center}
\begin{picture}(100,100)(0,0)
        \Line(0,70)(100,70) \Line(0,20)(100,20) 
        \Line(15,70)(15,20) \Line(85,70)(85,20)
        \Line(50,70)(85,45)
 \Text(32,73)[b]{1}
 \Text(65,73)[b]{2}
 \Text(88,60)[l]{3}
 \Text(88,30)[l]{4}
 \Text(71,53)[rt]{5}
\end{picture}
\caption{\label{triangle_ibp} The PentaBox represents a typical diagram where
       the triangle relation can be applied. 
In the present calculation, the lower left leg is massive.}
\end{center}
\end{figure}
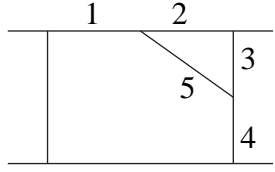

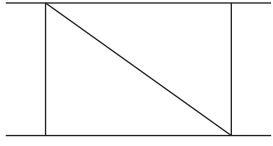
\begin{figure}
\begin{center}
\begin{picture}(100,100)(0,0)
       \Line(0,70)(100,70) \Line(0,20)(100,20) 
        \Line(15,70)(15,20) \Line(85,70)(85,20)
        \Line(15,70)(85,20)
\end{picture}
\caption{\label{cbox} The generic $\mbox{CBox}$. 
  For the $\mbox{CBox}_2$, the upper right external leg is massive.}
\end{center}
\end{figure}

At this stage, we are left with sets of scalar integrals of a given
topology in $4+2m-2\eps$-dimensions, and with raised powers of propagators,  
where $m$ is a non-negative integer.
Then, for some topologies like the PentaBox in fig. (\ref{triangle_ibp}), 
immediate simplifications are possible. By means of 
integration-by-parts \cite{'tHooft:1972fi,Chetyrkin:1981qh} 
these topologies reduce to simpler ones. 
For the PentaBox in fig. (\ref{triangle_ibp}), for instance, partial integration provides 
the following triangle relations, 
\begin{eqnarray}
\left[ \left( \dim  - 2 \nu_2 - \nu_3 - \nu_5 \right)
       - \nu_3 {\bf 3}^+ {\bf 2}^- 
       - \nu_5 {\bf 5}^+ \left( {\bf 2}^- - {\bf 1}^- \right) 
\right] \mbox{PentaBox}(m,\nu_1,\dots,\nu_7) & = & 0,
 \nonumber \\
\left[ \left( \dim  - \nu_2 - 2 \nu_3 - \nu_5 \right)
       - \nu_2 {\bf 2}^+ {\bf 3}^- 
       - \nu_5 {\bf 5}^+ \left( {\bf 3}^- - {\bf 4}^- \right) 
\right] \mbox{PentaBox}(m,\nu_1,\dots,\nu_7) & = & 0.
\end{eqnarray}
These relations can be used to eliminate the propagators $1$ and $4$, such that 
the PentaBox becomes reducible to the $\mbox{CBox}_2$ shown in fig. (\ref{cbox}).

After having performed obvious simplifications based on triangle relations, 
we are then able, to calculate directly all necessary scalar integrals with 
the method of nested sums \cite{Moch:2001zr}. For each topology, this requires 
analytical solutions valid in any dimension and for any (not necessarily
integer) power of the propagators in terms of nested sums. 
For the fermionic contributions to $e^+e^-\to q \qb g$ 
up to two loops, it suffices to have these analytical expressions for the $\mbox{CBox}_2$
\cite{Moch:2001zr},  the one-loop box with one external mass \cite{Anastasiou:1999cx}
and the one-loop triangle with two external masses \cite{Anastasiou:1999ui}.

We give the explicit representations as nested sums for the basic integrals.
As a short-hand notation, we use $\nu_{ij}=\nu_i+\nu_j$ for sums of powers of 
propagators in the following.
The one-loop triangle with two external masses is defined by
\begin{eqnarray}
     \mbox{Tri}_2(m,\nu_1,\nu_2,\nu_3;x_1) = 
         \left( -s_{123} \right)^{-m+\varepsilon+\nu_{123}}
         \int \frac{d^\dim k_1}{i \pi^{\dim /2}}
         \frac{1}{\left(-k_1^2\right)^{\nu_1}}
         \frac{1}{\left(-k_2^2\right)^{\nu_2}}
         \frac{1}{\left(-k_3^2\right)^{\nu_3}},
\end{eqnarray}
where $k_2=k_1-p_1-p_2$, $k_3=k_2-p_3$. It can be written as a 
combination of hypergeometric functions $_2F_1$.
The series representation for this integral is given by
\begin{eqnarray}
\lefteqn{
         \mbox{Tri}_2(m,\nu_1,\nu_2,\nu_3;x_1) \,= \,
             \frac{ 
                     \Gamma(\varepsilon-m+\nu_{23})
                     \Gamma(1-\varepsilon+m-\nu_{23})
                     \Gamma(m-\varepsilon-\nu_{13})
                  }{
                     \Gamma(\nu_1) \Gamma(\nu_2) \Gamma(\nu_3) 
                     \Gamma(2m-2\varepsilon-\nu_{123})
                   } 
}
\\ & & 
             \times 
             \sum\limits_{i=0}^\infty
                 \frac{x_1^{i}}{i!} 
             \left[
             x_1^{m-\varepsilon-\nu_{23}} 
             \frac{
                     \Gamma(i_1+\nu_1) 
                     \Gamma(i_1-\varepsilon+m-\nu_2) 
                  }{
                     \Gamma(i_1+1+m-\varepsilon-\nu_{23}) 
                   }
                 - 
             \frac{
                     \Gamma(i_1+\nu_3) 
                     \Gamma(i_1-m+\varepsilon+\nu_{123})
                  }{
                     \Gamma(i_1+1-m+\varepsilon+\nu_{23}) 
                   }
             \right]\, .
\nonumber 
\end{eqnarray}
The one-loop box integral is defined by 
\begin{eqnarray}
\lefteqn{
    \mbox{Box}(m,\nu_1,\nu_2,\nu_3,\nu_4;x_1,x_2) \,= }
\\ & &
        \left( -s_{123} \right)^{-m+\varepsilon+\nu_{1234}}
        \int \frac{d^\dim k_1}{i \pi^{\dim /2}}
         \frac{1}{\left(-k_1^2\right)^{\nu_1}}
         \frac{1}{\left(-k_2^2\right)^{\nu_2}}
         \frac{1}{\left(-k_3^2\right)^{\nu_3}}
         \frac{1}{\left(-k_4^2\right)^{\nu_4}},
\nonumber
\end{eqnarray}
where $k_2=k_1-p_1$, $k_3=k_2-p_2$ and $k_4=k_3-p_3$.
This integral can be expressed in terms of a combination of
Appell functions of the second kind 
and the series 
representation is given by:
\begin{eqnarray}
\lefteqn{
         \mbox{Box}(m,\nu_1,\nu_2,\nu_3,\nu_4;x_1,x_2) \,= }
\nonumber \\ & &
             \frac{ 
                     \Gamma(m-\varepsilon-\nu_{123})
                     \Gamma(m-\varepsilon-\nu_{234})
                     \Gamma(1+\nu_{123}-m+\varepsilon)
                     \Gamma(1+\nu_{234}-m+\varepsilon)
                  }{
                     \Gamma(\nu_1) \Gamma(\nu_2) \Gamma(\nu_3) \Gamma(\nu_4) 
                     \Gamma(2m-2\varepsilon-\nu_{1234})
                   } 
\nonumber \\ & &
             \times
             \sum\limits_{i_1=0}^\infty
             \sum\limits_{i_2=0}^\infty
                 \frac{x_1^{i_1}}{i_1!}
                 \frac{x_2^{i_2}}{i_2!} 
             \left[
             \frac{
                     \Gamma(i_1+\nu_3) 
                     \Gamma(i_2+\nu_2) 
                     \Gamma(i_1+i_2-m+\varepsilon+\nu_{1234})
                  }{
                     \Gamma(i_1+1-m+\varepsilon+\nu_{123}) 
                     \Gamma(i_2+1-m+\varepsilon+\nu_{234})
                   }
             \right.
\nonumber \\ & & 
             \mbox{} - x_1^{m-\varepsilon-\nu_{123}} 
             \frac{
                     \Gamma(i_1+m-\varepsilon-\nu_{12}) 
                     \Gamma(i_2+\nu_2) 
                     \Gamma(i_1+i_2+\nu_4) 
                  }{
                     \Gamma(i_1+1+m-\varepsilon-\nu_{123}) 
                     \Gamma(i_2+1-m+\varepsilon+\nu_{234}) 
                   }
\nonumber \\ & & 
             \mbox{} - x_2^{m-\varepsilon-\nu_{234}}
             \frac{
                      \Gamma(i_1+\nu_3) 
                      \Gamma(i_2+m-\varepsilon-\nu_{34}) 
                      \Gamma(i_1+i_2+\nu_1)
                   }{
                      \Gamma(i_1+1-m+\varepsilon+\nu_{123}) 
                      \Gamma(i_2+1+m-\varepsilon-\nu_{234}) 
                     }
\\ & & 
             \left. 
             + x_1^{m-\varepsilon-\nu_{123}} x_2^{m-\varepsilon-\nu_{234}}
             \frac{
                      \Gamma(i_1+m-\varepsilon-\nu_{12}) 
                      \Gamma(i_2+m-\varepsilon-\nu_{34}) 
                      \Gamma(i_1+i_2+m-\varepsilon-\nu_{23})
                   }{
                      \Gamma(i_1+1+m-\varepsilon-\nu_{123}) 
                      \Gamma(i_2+1+m-\varepsilon-\nu_{234}) 
                     } 
             \right].
\nonumber 
\end{eqnarray}
Finally, the two-loop $\mbox{CBox}_2$ is defined by
\begin{eqnarray}
\lefteqn{
     \mbox{CBox}_2(m,\nu_1,\nu_2,\nu_3,\nu_4,\nu_5;x_1,x_2) \,=} 
\\ & & 
         \left( -s_{123} \right)^{-2m+2\varepsilon+\nu_{12345}}
         \int \frac{d^\dim k_1}{i \pi^{\dim /2}}
         \int \frac{d^\dim l_5}{i \pi^{\dim /2}}
         \frac{1}{\left(-k_1^2\right)^{\nu_1}}
         \frac{1}{\left(-l_2^2\right)^{\nu_2}}
         \frac{1}{\left(-l_3^2\right)^{\nu_3}}
         \frac{1}{\left(-k_4^2\right)^{\nu_4}}
         \frac{1}{\left(-l_5^2\right)^{\nu_5}},
\nonumber 
\end{eqnarray}
with $l_2 = k_1 + l_5 - p_1$, $l_3 = l_2 - p_2$, $k_4 = k_1 - p_{123}$.
The formula for this integral is given by:
\begin{eqnarray}
\lefteqn{
         \mbox{CBox}_2(m,\nu_1,\nu_2,\nu_3,\nu_4,\nu_5;x_1,x_2) \,=} 
\nonumber \\ & & 
             \frac{ 
                     \Gamma(2m-2\varepsilon-\nu_{1235})
                     \Gamma(1+\nu_{1235}-2m+2\varepsilon)
                     \Gamma(2m-2\varepsilon-\nu_{2345})
                     \Gamma(1+\nu_{2345}-2m+2\varepsilon)
                  }{
                     \Gamma(\nu_1) \Gamma(\nu_2) \Gamma(\nu_3) \Gamma(\nu_4) \Gamma(\nu_5)
                     \Gamma(3m-3\varepsilon-\nu_{12345})
                   } 
\nonumber \\ & & 
             \times
             \frac{ 
                     \Gamma(m-\varepsilon-\nu_5) 
                     \Gamma(m-\varepsilon-\nu_{23})
                  }{
                     \Gamma(2m-2\varepsilon-\nu_{235})
                   }
             \sum\limits_{i_1=0}^\infty
             \sum\limits_{i_2=0}^\infty
                 \frac{x_1^{i_1}}{i_1!}
                 \frac{x_2^{i_2}}{i_2!} 
\nonumber \\ & &
             \times \left[
             \frac{
                     \Gamma(i_1+\nu_3) 
                     \Gamma(i_2+\nu_2) 
                     \Gamma(i_1+i_2-2m+2\varepsilon+\nu_{12345})
                     \Gamma(i_1+i_2-m+\varepsilon+\nu_{235})
                  }{
                     \Gamma(i_1+1-2m+2\varepsilon+\nu_{1235}) 
                     \Gamma(i_2+1-2m+2\varepsilon+\nu_{2345})
                     \Gamma(i_1+i_2+\nu_{23})
                   }
             \right.
\nonumber \\ & &
             \mbox{} - x_1^{2m-2\varepsilon-\nu_{1235}} 
\nonumber \\ & &
             \times
             \frac{
                     \Gamma(i_1+2m-2\varepsilon-\nu_{125}) 
                     \Gamma(i_2+\nu_2) 
                     \Gamma(i_1+i_2+\nu_4) 
                     \Gamma(i_1+i_2+m-\varepsilon-\nu_1)
                  }{
                     \Gamma(i_1+1+2m-2\varepsilon-\nu_{1235}) 
                     \Gamma(i_2+1-2m+2\varepsilon+\nu_{2345}) 
                     \Gamma(i_1+i_2+2m-2\varepsilon-\nu_{15})
                   }
\nonumber \\ & &
             \mbox{} - x_2^{2m-2\varepsilon-\nu_{2345}}
\nonumber \\ & &
             \times
             \frac{
                      \Gamma(i_1+\nu_3) 
                      \Gamma(i_2+2m-2\varepsilon-\nu_{345}) 
                      \Gamma(i_1+i_2+\nu_1)
                      \Gamma(i_1+i_2+m-\varepsilon-\nu_4)
                   }{
                      \Gamma(i_1+1-2m+2\varepsilon+\nu_{1235}) 
                      \Gamma(i_2+1+2m-2\varepsilon-\nu_{2345}) 
                      \Gamma(i_1+i_2+2m-2\varepsilon-\nu_{45})
                     }
\nonumber \\ & &
             + x_1^{2m-2\varepsilon-\nu_{1235}} x_2^{2m-2\varepsilon-\nu_{2345}}
             \frac{
                      \Gamma(i_1+2m-2\varepsilon-\nu_{125}) 
                      \Gamma(i_2+2m-2\varepsilon-\nu_{345}) 
                   }{
                      \Gamma(i_1+1+2m-2\varepsilon-\nu_{1235}) 
                      \Gamma(i_2+1+2m-2\varepsilon-\nu_{2345}) 
                     } 
\nonumber \\ & &
             \left. \times
             \frac{
                      \Gamma(i_1+i_2+2m-2\varepsilon-\nu_{235})
                      \Gamma(i_1+i_2+3m-3\varepsilon-\nu_{12345})
                   }{
                      \Gamma(i_1+i_2+4m-4\varepsilon-\nu_{12345}-\nu_5)
                    }
             \right].
\end{eqnarray}

Then, the evaluation of the multiple nested sums proceeds systematically with the
help of the algorithms of \cite{Moch:2001zr}. These algorithms rely on the
algebraic properties of the so called $Z$-sums and allow to solve by means
of recursion the nested sums in terms of a given basis in $Z$-sums 
to any order in the expansion parameter $\eps$.
$Z$-sums are defined by
\bq 
\label{defZsums}
  Z(n;m_1,...,m_k;x_1,...,x_k) & = & \sum\limits_{n\ge i_1>i_2>\ldots>i_k>0}
     \frac{x_1^{i_1}}{{i_1}^{m_1}}\ldots \frac{x_k^{i_k}}{{i_k}^{m_k}}\, .
\eq
They are
generalizations of Euler-Zagier sums \cite{Euler,Zagier} 
\bq
Z_{m_1,...,m_k}(n) & = & Z(n;m_1,...,m_k;1,...,1)\, ,
\eq
or of harmonic sums
%\cite{Gonzalez-Arroyo:1979df,Gonzalez-Arroyo:1980he,Vermaseren:1998uu,Blumlein:1998if}
\cite{Gonzalez-Arroyo:1979df} - \cite{Blumlein:1998if}
involving multiple ratios of scales.
The latter are well known from calculations of Mellin moments of
deep-inelastic structure functions
%\cite{Gonzalez-Arroyo:1979df,Gonzalez-Arroyo:1980he,Kazakov:1988jk,Kazakov:1990jm,Moch:1999eb,Vermaseren:2000we}.
\cite{Gonzalez-Arroyo:1979df}, \cite{Kazakov:1988jk} - \cite{Vermaseren:2000we}.
An important subset of $Z$-sums are multiple polylogarithms 
%\cite{Goncharov,Borwein,Remiddi:1999ew}, 
\cite{Goncharov}-\cite{Remiddi:1999ew}, 
obtained by letting $n$ go to infinity in eq. (\ref{defZsums}):
\bq
\label{multipolylog}
\mbox{Li}_{m_k,...,m_1}(x_k,...,x_1) & = & Z(\infty;m_1,...,m_k;x_1,...,x_k).
\eq

The key feature of $Z$-sums is the fact, 
that they interpolate between
Goncharov's multiple polylogarithms and Euler-Zagier sums, 
which occur in the expansion of $\Gamma$-functions,
\bq
\label{expgamma}
\lefteqn{
\Gamma(n+\eps) = \Gamma(1+\eps) \Gamma(n) } & & \nonumber \\
& & \times \left( 1 + \eps Z_1(n-1) + \eps^2 Z_{11}(n-1) + \eps^3 Z_{111}(n-1) + ... 
+ \eps^{n-1} Z_{11...1}(n-1) \right). 
\eq

In summary, the evaluation of the multiple nested sums proceeds by expanding
the $\Gamma$-functions to the desired order, and by using then the algebraic properties
of the $Z$-sums. 
In this way, we could calculate all loop integrals contributing to the one-
and two-loop virtual amplitudes very efficiently in terms of multiple polylogarithms.
All algorithms for this procedure have been implemented on a computer in 
symbolic manipulation programs. 
We have chosen to work with FORM~\cite{Vermaseren:2000nd} 
and in the GiNaC framework \cite{Bauer:2000cp,Weinzierl:2002hv}.

\section{Results}
\label{sec:results}
\def\R{{\rm R}}
\def\Li{{\rm Li}}
\def\Li(#1,#2){{{\rm Li}_{#1}(#2)}}
\def\Lii(#1,#2,#3,#4){{{\rm Li}_{#1,#2}\left(#3,#4\right)}}
\def\Liii(#1,#2,#3,#4,#5,#6){{{\rm Li}_{#1,#2,#3}\left(#4,#5,#6\right)}}
\def\z#1{{\zeta_{#1}}}
\def\RR#1{{\rm R_{#1}}}

Let us now discuss our results for the finite coefficients $c_{i}^{(1),\fin}$ 
through order $\epsilon^2$ and the $n_f$-contributions to $c_{i}^{(2),\fin}$. 

We have made the following checks on our result. 
First of all, after UV renormalization, the infrared poles agree with the 
structure predicted by Catani \cite{Catani:1998bh}. 
This provides a strong check of the complete pole structure of our result.
Secondly, having calculated all coefficients $c_1$--$c_{13}$, 
we could use the various relations eqs. (\ref{eq:constraint1}) and (\ref{eq:constraint2}) 
between $c_1$--$c_{13}$ as a cross-check.
Note that for $c_6$ there is an additional symmetry: one can show
that the combination $\x1\* c_6$ is symmetric under the 
exchange $(\x1\leftrightarrow\x2)$, 
\begin{eqnarray}
\label{eq:c6nnsym}
\x1 c_{6}(\x1,\x2)  &=& \x2  c_{6}(\x2,\x1) \, . 
\end{eqnarray}
Finally, we could compare with recent work of Garland et al.. 
They have obtained by an independent method the result for the squared 
matrix elements \cite{Garland:2001tf}, 
i.e. the interference of the two-loop amplitude with the Born amplitude, 
and the interference of the one-loop amplitude with itself, 
as well as the result for the one- and two-loop amplitude \cite{Garland:2002ak}. 
Their results are given in terms of one- and two-dimensional harmonic 
polylogarithms, which form a subset of the multiple polylogarithms 
%\cite{Goncharov,Borwein,Remiddi:1999ew}. 
\cite{Goncharov}-\cite{Remiddi:1999ew}. 
Thus, we have performed the 
comparision analytically. We agree with both of their results. 

Here, we present explicitly the function $c_{i}^{(2),\fin}$, $i=2,4,6,12$. 
All formulae for the one-loop amplitude are deferred 
to appendix \ref{sec:appendixresults}.

\begin{eqnarray}
  &&c_2^{(2),\fin}(\x1,\x2) =
%%START
%%c_2=
    \nf\*N\*\bigg(
    -2\*{\x2\over (\x1+\x2)^3}\*\RR1(\x1,\x2)
    +{\RR2(\x1,\x2) \over (1-\x1-\x2)^2}
  -{3 \over 4} \*{\ln(\x1)^2\over (1-\x1)\*(1-\x1-\x2)}
  \nn\\&&
  +{1 \over 6} \*{(-6+19\*\x2+25\*\x1)\over(\x1+\x2)\*(1-\x1-\x2)\*(1-\x1)}
  \*\ln(\x1)
  +{1 \over 12}\*{(-2\*\x1+\x1\*\x2+\x1^2-2+2\*\x2)\over
    \x1\*(1-\x1)\*(1-\x2)\*(1-\x1-\x2)}\*\z2
  \nn\\&&
  +{1 \over 4} \*{(\ln(\x1)+\ln(\x2))\*\z2\over (1-\x1-\x2)^2}
  -{1 \over 4} \*{(2-2\*\x2+\x1)\over\x1\*(1-\x1-\x2)\*(1-\x2)}\*\ln(\x2)^2
  \nn\\&&
  +{1\over 18}\*{(44\*\x1-31\*\x1\*\x2+13\*\x1^2+26\*\x2-26\*\x2^2)
    \over \x1\*(\x1+\x2)\*(1-\x2)\*(1-\x1-\x2)}\*\ln(\x2)
  +{1 \over 12} \*
  \bigg[-8-{(1+\x1)\over (1-\x2)}
  -{24\*\x1\over (\x1+\x2)}
  \nn\\&&
  +2\*{(13+6\*\x1)\over (1-\x1-\x2)}-10\*{\x2\over \x1}
  -3\*{(1+\x2)\over (1-\x1)}\bigg]
  \*{\R(\x1,\x2)\over (\x1+\x2)\*(1-\x1-\x2)}
  +{1 \over 12} \*
  \bigg[-{12\over \x1\*(\x1+\x2)}
  \nn\\&&
  +{24\over (\x1+\x2)^2}
  -{1\over \x1\*(1-\x2)}
  +3\*{(1+4\*\x1)\over \x1\*(1-\x1)\*(1-\x1-\x2)}\bigg]
  \*\big[\Li(2,1-\x1)-\Li(2,1-\x2)\big]
  \nn\\&&
  +{1 \over 2} \*{1\over \x1}\*
  \bigg[-{1\over (1-\x2)}+3\*{1\over (1-\x1)\*(1-\x1-\x2)}\bigg]
  \*\Li(2,1-\x2)
  \bigg)
  \nn\\&&
  + {\nf\over N}\*\bigg(
  {(\x1^2-\x2^2)\over 3\*\x1^2\*\x2^2}\*\bigg[
  \Li(3,\x1+\x2)
  +\Li(3,1-\x1-\x2)
  +{1 \over 2}\*\ln(\x1+\x2)\*\ln(1-\x1-\x2)^2
  \nn\\&&
  -{1 \over 4}\*\ln(\x1)\*\R(\x1,1-\x1-\x2)
  +{1 \over 4}\*(\ln(\x1)^2-\ln(\x1)\*\ln(\x2)  +\ln(\x2)^2)\*\ln(1-\x1-\x2)
  \nn\\&&
  -{1 \over 4}\*\ln(\x2)\*\R(\x2,1-\x1-\x2)
  \bigg]
  +{1\over 6\*\x1\*\x2}\*\bigg[
   {1 \over 2} \*\Li(2,1-\x2)
  -{5 \over 2} \*\R(\x1,1-\x1-\x2)
  -2\*\z2
  \nn\\&&
  +{5 \over 2} \*\Li(2,1-\x1)
  -\Li(2,1-\x1-\x2)
  -\ln(1-\x1-\x2)^2
  -\ln(\x1+\x2)\*\ln(1-\x1-\x2)
  \nn\\&&
  -{1 \over 2} \*\R(\x2,1-\x1-\x2)
  \bigg]
  +{1\over 3\*\x2^2}\*\bigg[
   {1 \over 4}\*\ln(\x1)^2\*\ln(1-\x1-\x2)
  -{1 \over 2} \*\ln(\x1)\*\ln(1-\x1-\x2)^2
  \nn\\&&
  -{1 \over 2} \*\ln(\x1)^2\*\ln(1-\x1)
  -{1 \over 4} \*\ln(\x2)\*(\Li(2,1-\x1)-\Li(2,1-\x2))
  -\Li(3,\x1)
  -\Li(3,1-\x1)
  \nn\\&&
  -\ln(\x1)\*\R(\x1,1-\x1-\x2)
  +{19 \over 6} \*\R(\x1,1-\x1-\x2)
  \bigg]
  +{1\over 3\*\x1^2}\*\bigg[
  {1 \over 2}  \*\ln(\x2)\*\ln(1-\x1-\x2)^2
  \nn\\&&
  -{1 \over 4} \*\ln(\x2)^2\*\ln(1-\x1-\x2)
  +{1 \over 2}  \*\ln(\x2)^2\*\ln(1-\x2)
  -{1 \over 4} \*\ln(\x1)\*(\Li(2,1-\x1)-\Li(2,1-\x2))
  \nn\\&&
  +\Li(3,\x2)
  +\Li(3,1-\x2)
  +\ln(\x2)\*\R(\x2,1-\x1-\x2)
  -{19 \over 6} \*\R(\x2,1-\x1-\x2)
  \bigg]
  \nn\\&&
  +{1 \over 18} \*{(19\*\x1+13\*\x2)\over \x1\*\x2\*(\x1+\x2)}\*\ln(1-\x1-\x2)
  +{1 \over 4} \*{(-1+3\*\x2+\x1)\over\x2\*(1-\x1)\*(1-\x1-\x2)}\*\ln(\x1)^2
  \nn\\&&
  -{1\over 18}\*{(-19+19\*\x1+51\*\x2)\over
  \x2\*(1-\x1)\*(1-\x1-\x2)}\*\ln(\x1)
  +{1 \over 3} \* {(1+\x1+\x2) \over \x1\*(1-\x1-\x2)\*(\x1+\x2)}\*\R(\x2,1-\x1-\x2)
  \nn\\&&
  +{1 \over 4} \*{(1-\x2+\x1)\over \x1\*(1-\x2)\*(1-\x1-\x2)}\*\ln(\x2)^2
  -{1 \over 18}\*{(19-19\*\x2+13\*\x1)
    \over\x1\*(1-\x2)\*(1-\x1-\x2)}\*\ln(\x2)
  +{1 \over 3}\*{\Li(2,1-\x2)\over \x1\*(1-\x2)}
  \nn\\&&
  -{1 \over 12} \*\bigg[
  {1\over \x1\*(1-\x2)}+{1\over \x2\*(1-\x1)}
  -{2\over \x1\*(1-\x1)\*(1-\x1-\x2)}\bigg]\*
  (\z2+\R(\x1,\x2)
  \nn\\&&
  -\Li(2,1-\x1)-\Li(2,1-\x2))
  +{1 \over 3}\*{ (\Li(2,1-\x1)-\Li(2,1-\x2)-\R(\x1,1-\x1-\x2)) \over \x1\*(\x1+\x2)}
  \nn\\&&
  -{\Li(2,1-\x1) \over (1-\x1)\*(1-\x1-\x2)}
  +{1 \over 3}\*
  {(\Li(2,1-\x1)+2\*\R(\x1,1-\x1-\x2)) \over \x2\*(1-\x1-\x2)} 
  -{2 \over 3}\*{\Li(2,1-\x2) \over \x1\*(1-\x1-\x2)}
  \bigg)
  \nn\\&&
  +i\*\pi\*\nf\*N\*\bigg(
  {3 \over 2} \*{\ln(\x1)\over (1-\x1)\*(1-\x1-\x2)}
  +{1 \over 2} \*{(2-2\*\x2+\x1)\over \x1\*(1-\x2)\*(1-\x1-\x2)}\*\ln(\x2)
  +{3 \over 2} \*{\R(\x1,\x2) \over (1-\x1-\x2)^2}\bigg) 
  \nn\\&&
  +{i\*\pi\*\nf \over N}\*\bigg(
   {1 \over 2} \*{1\over \x2^2}\*\R(\x1,1-\x1-\x2)
  -{1 \over 2} \*{1\over \x1^2}\*\R(\x2,1-\x1-\x2)
  +{1 \over 2} \*{\ln(1-\x1-\x2)\over \x1\*\x2}
  \nn\\&&
  -{1 \over 2} \*{(-1+3\*\x2+\x1)\over \x2\*(1-\x1)\*(1-\x1-\x2)}\*\ln(\x1)
  -{1 \over 2} \*{(1-\x2+\x1)\over\x1\*(1-\x2)\*(1-\x1-\x2)}\*\ln(\x2)
  \bigg)
  %%
  %%
%%:
%%STOP
,\label{eq:c2nn}
\end{eqnarray}

\begin{eqnarray}
  &&c_4^{(2),\fin}(\x1,\x2) =
%%START
%%c_4=
 \nf\*N\*\bigg(
   {1 \over 36} \*{\ln(\x2) \over (1-\x2)}\*
  \bigg[36\*{(1+\x1)\over (\x1+\x2)}-{59\over \x1}-{50\over \x1\*(1-\x2)}
  +12\*{(1-3\*\x1)\*(1-\x1)\over \x1\*(1-\x1-\x2)}
  \bigg]
    \nn\\&&
  -{19 \over 18} \*{1\over \x1\*(1-\x2)}
  +{(\x1-\x2)\over (\x1+\x2)^3}\*\RR1(\x1,\x2)
 -\bigg[{5 \over 12} \*{(2-\x2)\over \x1\*(1-\x2)^2}
  +{2\over (\x1+\x2)^2}
  \bigg]\*\Li(2,1-\x2)
  \nn\\&&
 +{1 \over 4} \*{(2-\x2)\over \x1\*(1-\x2)^2}\*\bigg[
 \ln(\x2)^2+{1 \over 3} \*\z2\bigg]
  -{1 \over 12} \*{(7\*\x1-5\*\x2-6\*\x1\*\x2+5\*\x2^2+\x1^2)
    \over (1-\x2)\*\x1\*(1-\x1-\x2)\*(\x1+\x2)}\*\ln(\x1)
  \nn\\&&
  -{\R(\x1,\x2)\over (1-\x1-\x2)\*(1-\x2)\*(\x1+\x2)}\*
  \bigg[{17 \over 6 }\*\x2-{35 \over 12} \*\x1
  -{19 \over 12} +{1 \over 12} \*{(1+\x1)\over (1-\x2)}
  -{1 \over 3} \*{(1-3\*\x1)\*(1-\x1)\over (1-\x1-\x2)}
  \nn\\&&
  +{5\*\x2^2 \over 12\*\x1} -{5 \over 6} \*{\x2\over \x1}
  +2\*{\x1\*(1+\x1)\over (\x1+\x2)}\bigg]
  -\bigg[
 {1 \over 12} \*{(2-\x2)\over \x1\*(1-\x2)^2}
  -{2\over (\x1+\x2)^2}
  \bigg]
  \*\Li(2,1-\x1)
  \bigg)
  \nn\\&&
  +{\nf\over N}\*\bigg(    
  -{1 \over 3} \*{1\over \x1^2}\*\bigg[\Li(2,1-\x1-\x2)
  +\ln(1-\x1-\x2)^2
  +\ln(\x1+\x2)\*\ln(1-\x1-\x2)
  \bigg]
    \nn\\&&
  +{19 \over 18} \*{1\over (1-\x2)\*\x1}
  +{(\x1+\x2)\over 3\*\x1^3}\*\bigg[
  {1 \over 2}\*\ln(\x1)^2 \*\ln(1-\x1-\x2)
  +(\ln(\x1+\x2)-\ln(\x2))\*\ln(1-\x1-\x2)^2
  \nn\\&&
  +{1 \over 2} \*\ln(\x1)\*\Li(2,1-\x1)
  -{1 \over 2} \*\Li(2,1-\x2)\*\ln(\x1)
  -2\*\Li(3,1-\x2)
  +2 \*\Li(3,\x1+\x2)
  \nn\\&&
  +2 \*\Li(3,1-\x1-\x2)
  -2\*\Li(3,\x2)
  -{1 \over 2} \*\ln(\x2)\*\ln(\x1)\*\ln(1-\x1-\x2)
  +\ln(\x2)^2\*\ln(1-\x1-\x2)
  \nn\\&&
  -{1 \over 2} \*\ln(\x1)\*\R(\x1,1-\x1-\x2)
  -{5 \over 2} \*\ln(\x2)\*\R(\x2,1-\x1-\x2)
  -\ln(\x2)^2\*\ln(1-\x2)\bigg]
  \nn\\&&
  +{1 \over 36} \*{(76\*\x2\*(1-\x2)+61\*\x1-11\*\x1\*\x2)\over
    \x1^2\*(1-\x2)^2}\*\ln(\x2)
  +{(17\*\x1^2+61\*\x1\*\x2+38\*\x2^2)\over 18\*\x1^3\*(\x1+\x2)}
  \*\R(\x2,1-\x1-\x2)
  \nn\\&&
  +{1 \over 4} \*{(-2\*\x2+2\*\x2^2-2\*\x1+\x1\*\x2)\over\x1^2\*(1-\x2)^2}
  \*\bigg[{1 \over 3} \*\R(\x1,\x2)+\ln(\x2)^2\bigg]
  +{(\x1-\x2)\over 6\*\x1^2\*(\x1+\x2)}\*\R(\x1,1-\x1-\x2)
  \nn\\&&
  +{(-8+14\*\x2-6\*\x2^2-2\*\x1+\x1\*\x2)\over 12\*\x1^2\*(1-\x2)^2}
  \*\z2
  -{1 \over 12} \*{\ln(\x1)\over \x1\*(1-\x2)}
  +{1 \over 9} \*{(16\*\x1+19\*\x2)\over\x1^2\*(\x1+\x2)}\*\ln(1-\x1-\x2)
  \nn\\&&
  -{1 \over 12} \*{(-10\*\x1^2+5\*\x1^2\*\x2-14\*\x1+8\*\x1\*\x2
    +\x1\*\x2^2-10\*\x2+10\*\x2^2)\over (\x1+\x2)\*\x1^2\*(1-\x2)^2}
  \*\Li(2,1-\x2)
  \nn\\&&
  -{1 \over 12} \*{(-2\*\x1^2+\x1^2\*\x2+2\*\x1-8\*\x1\*\x2
    +5\*\x1\*\x2^2-2\*\x2+2\*\x2^2)\over (\x1+\x2)\*\x1^2\*(1-\x2)^2}
  \*\Li(2,1-\x1)
  \bigg)
  \nn\\&&
   +{i\*\pi\*\nf \over N}\*\bigg(
  {1\over \x1^2}\*\ln(1-\x1-\x2)
  -{1 \over 2} \*{(-2\*\x2+2\*\x2^2-2\*\x1+\x1\*\x2)\over\x1^2\*(1-\x2)^2}
  \*\ln(\x2)
  +{1 \over 2} \*{1\over \x1\*(1-\x2)}
  \nn\\&&
  +{(\x1+\x2)\over \x1^3 }\*\R(\x2,1-\x1-\x2)
  \bigg)
  -i\*\pi\*\nf\*N\*\bigg(
  {1 \over 2} \*{(2-\x2)\over\x1\*(1-\x2)^2} \*\ln(\x2)
  +{1 \over 2} \*{1\over \x1\*(1-\x2)}\bigg) %%:
%%STOP
,\label{eq:c4nn}
\end{eqnarray}

\begin{eqnarray}
  &&\x1c_6^{(2),\fin}(\x1,\x2) =
%%START
%%\x1c_6=
  {\nf \over N}\*\bigg(
   {1\over 2}\*\bigg[{1\over 3} \*\ln(1-\x1 -\x2 )
   -{91 \over 108} 
   +{1 \over 4} \*\z2 
   +{1 \over 3} \*\ln(\x1 +\x2 )
   \bigg]\*\ln(1-\x1 -\x2 )
  \nn\\&&
  +{1 \over 36} \*{(52\*\x1^2\*\x2 +47\*\x1 \*\x2^2-45\*\x1 \*\x2 
    -14\*\x2^2+14\*\x2^3-19\*\x1^2
    +19\*\x1^3)\over (1-\x1 -\x2 ) \*(\x1 +\x2)\*\x2 }
  \*\R(\x1 ,1-\x1 -\x2 )
  \nn\\&&
  +{(\x1+2\*\x2)\over 6\*\x2}\* \bigg[
  +{1 \over 2} \*\ln(1-\x1)\*\ln(\x1)^2
  -{1 \over 4} \*\ln(\x2)\*\Li(2,1-\x2)
  +{1 \over 4} \*\Li(2,1-\x1)\*\ln(\x2)
  \nn\\&&
  +{1 \over 2} \*
  \ln(\x1 )\*\ln(1-\x1 -\x2 )^2
  +\Li(3,\x1)
  +\Li(3,1-\x1)
  \bigg]
  +{1 \over 6} \*\Li(2,1-\x1 -\x2 )
  \nn\\&&
  +{(\x2^2+\x1^2+4\*\x1 \*\x2 )\over 12\*\x1\*\x2 }\*\bigg[
  +{1 \over 4} \*\ln(\x2 )\*\ln(\x1 )\*\ln(1-\x1 -\x2 )
  -{1 \over 2} \*\ln(\x1 +\x2 )\*\ln(1-\x1 -\x2 )^2
  \nn\\&&
  -\Li(3,1-\x1 -\x2)
  -\Li(3,\x1 +\x2 )
  \bigg]
  +{1 \over 24} \*{(\x2^2+5\*\x1^2
  +12\*\x1 \*\x2 )\over \x1\*\x2 }\*\ln(\x1 )\*\R(\x1 ,1-\x1 -\x2 )
  \nn\\&&
  +{1 \over 36} \*{(-43\*\x1 +31\*\x1^2+63\*\x1 \*\x2 +12
    -12\*\x2 )\over (1-\x1 ) \*(1-\x1 -\x2 )} \*\ln(\x1 )
  +{1 \over 8} \*{\x1\*(1-\x1 -3\*\x2 )\over 
    (1-\x1 )\*(1-\x1 -\x2)}\*\ln(\x1 )^2
  \nn\\&&
  -{1 \over 24} \*{(2\*\x1^2+\x2^2+6\*\x1 \*\x2 )\over \x1\*\x2}
  \*\ln(\x1 )^2\*\ln(1-\x1 -\x2 ) 
  -{4085 \over 2592} 
  +{1 \over 72} \*\z3 
  \nn\\&&
  -{1 \over 24} \*{(\x1 +2\*\x1 \*\x2 
  -1+\x2 )\over (1-\x2 )\*(1-\x1 -\x2 ) }\*\z2 
  -{1 \over 24} \*{\x1\*(-1+4\*\x2)\over (1-\x2 )\*(1-\x1 -\x2 )}
  \*\R(\x1 ,\x2 )
  \nn\\&&
  +{1 \over 24} \*
  \bigg[-{3\over (1-\x2)}-{15\over (1-\x1)}
  -{2\*(4\*\x2-5)\over (1-\x1-\x2)}
  -{8\*\x2\over (\x1+\x2)}\bigg]
  \*\Li(2,1-\x1 )
  \bigg)
  \nn\\&&
  +\nf \*N\*\bigg(
  -{1 \over 72} \*{(-82\*\x1 +41\*\x1^2+14\*\x1 \*\x2 
  +41-41\*\x2 )\over (1-\x1 )\*(1-\x1 -\x2 )}\*\ln(\x1 )^2
  -{7 \over 36} \*\z3   +{4345 \over 2592}
  \nn\\ &&
  +{1 \over 36} \*{(31-56\*\x1 -31\*\x2 +25\*\x1^2-20\*\x1 \*\x2 )
    \over (1-\x1 ) \*(1-\x1 -\x2 )}
  \*\ln(\x1 )
  +{1 \over 72} \*
  \bigg[-33\*\x1^2+58\*\x1+58
  -{58\over (1-\x1)}
  \nn\\ &&
  +{(9\*\x1^2+9\*\x1+58)\over (1-\x2)}
  -{(4\*\x1^3-51\*\x1^2+58-69\*\x1)
    \over (1-\x1-\x2)}\bigg]
  \*{\R(\x1 ,\x2 )\over (1-\x1 -\x2 )\*(\x1 +\x2)}
  \nn\\ &&
  -{1 \over 12} \*{(2\*\x1^2+\x1 \*\x2 +2\*\x2^2)\over (\x1 +\x2 )^2} 
  \*\RR1(\x1 ,\x2 )
  +{1 \over 12} \*{(2-4\*\x1 -4\*\x2 +2\*\x1^2+\x1 \*\x2 +2\*\x2^2)
    \over (1-\x1 -\x2 )^2}
  \*\RR2(\x1 ,\x2 )
  \nn\\ &&
  +{1 \over 72} \*\bigg[11+{9\over (1-\x2)}
  +{45\over (1-\x1)}-{9\*(4\*\x1+1)\over (1-\x1-\x2)}
  -{36\*\x1\over (\x1+\x2)}\bigg]
  \*\Li(2,1-\x1 )
  \nn\\ &&
  -{1 \over 8} \*{\x1\*\x2 \*\ln(\x1 )\over (1-\x1 -\x2 )^2} \*\z2 
  -{1 \over 144} \*{(1340\*\x1 -1023\*\x1 \*\x2 -670\*\x1^2
    +688\*\x1^2\*\x2 -335)
    \over (1-\x1 )\*(1-\x2 ) \*(1-\x1 -\x2 ) }\*\z2
  \bigg)
  \nn\\&&
 + i\*\pi\*\nf \*N\*\bigg(
  {1 \over 24} \*{(-31+13\*\x1 )\*\x2 +31\*(1-\x1 )^2
    \over  (1-\x1 )\* (1-\x1 -\x2 ) }\*\ln(\x1 )
  +{1 \over 8} \*{
    \x1\*\x2-8\*\x1+2+4\*\x1^2
    \over  (1-\x1 -\x2 )^2 }\*\R(\x1 ,\x2 )
  \nn\\&&
  +{1 \over 12}  \*\z2 -{5 \over 72}  
  \bigg)
  +{i\*\pi\*\nf \over N}\*{1 \over 4} \*\bigg(
  {\x1\*(-1+\x1 +3\*\x2 )\over (1-\x1 )\*(1-\x1 -\x2 )}\*\ln(\x1 )
  -{1 \over \x1} \*(\x2 +2\*\x1 )\*\R(\x2 ,1-\x1 -\x2 )
  \nn\\ &&
  -2\*\ln(1-\x1 -\x2 )
  -{1 \over 2}\*\z2 
  -{191 \over 54} \bigg) %%:
%%STOP
  + (\x1\leftrightarrow\x2),
\label{eq:c6nn}
\end{eqnarray}

\begin{eqnarray}
  &&c_{12}^{(2),\fin}(\x1,\x2) =
%%START
%%c_{12}=
  \nf\*N\*\bigg(
  3\*{\ln(\x1)\over (\x1+\x2)^2}
  -{1\over 2} \*{1\over \x1\*(1-\x2)}\*\bigg[
  -{1 \over 2} \*\ln(\x2)^2
  -{1 \over 6} \*\z2+\Li(2,1-\x2)\bigg]
  \nn\\&&
  -{1 \over 18} \*{(13\*\x1^2+36\*\x1-10\*\x1\*\x2-18\*\x2+31\*\x2^2)
  \over (\x1+\x2)^2\*\x1\*(1-\x2)}
  \*\ln(\x2)
  +{(-\x2^2-2\*\x1+4\*\x2+\x1^2)\over (\x1+\x2)^4}\*\RR1(\x1,\x2)
  \nn\\&&
  -{1 \over 12} \*{1\over \x1\*(\x1+\x2)^2}\*\bigg[
  5\*\x2+42\*\x1+5-{(1+\x1)^2\over (1-\x2)}
  -4\*{(1-3\*\x1+3\*\x1^2)\over (1-\x1-\x2)}-72\*{\x1^2\over (\x1+\x2)}\bigg]
  \*\R(\x1,\x2)
  \nn\\&&
  +\bigg[
  {(1+2\*\x1)\over \x1\*(\x1+\x2)^2}
  -{6\over (\x1+\x2)^3}-{1 \over 12} \*{1\over \x1\*(1-\x2)}
  \bigg]\*(\Li(2,1-\x1)-\Li(2,1-\x2))
  -{1\over (\x1+\x2)\*\x1}
  \bigg)
  \nn\\&&
  \nn\\&&
  +{\nf\over N}\*\bigg(  -{19 \over 9} \*{1\over \x1\*(\x1+\x2)}
  +{1 \over 18} \*{(13\*\x1^2-29\*\x1+42\*\x1\*\x2-38\*\x2+38\*\x2^2)
  \over \x1^2\*(\x1+\x2)\*(1-\x2)} \*\ln(\x2)
  \nn\\&&
  +{1 \over 4} \*{(2-2\*\x2-\x1)\over \x1^2\*(1-\x2)}\*\ln(\x2)^2
  +{1 \over 9}\*{(-32\*\x1-19\*\x2+16\*\x1^2+35\*\x1\*\x2+19\*\x2^2)\over 
    \x1^2\*(\x1+\x2)^2} \*\ln(1-\x1-\x2)
  \nn\\&&
  -{(1-\x1-\x2)\over 3\*\x1^3}\*\bigg[
  \bigg\{
  \ln(\x2)^2
  -{1\over 2}\*\ln(\x1)\*\ln(\x2)
  +{1 \over 2} \*\ln(\x1)^2
  +\ln(\x1+\x2)\*\ln(1-\x1-\x2)
  \nn\\&&
  -\ln(\x2)\*\ln(1-\x1-\x2)
  \bigg\}\*\ln(1-\x1-\x2)
  -\ln(\x2)^2\*\ln(1-\x2)
  -{1 \over 2} \*\Li(2,1-\x2)\*\ln(\x1)
  \nn\\&&
  -2 \*\Li(3,1-\x2)
  -{2 } \*\Li(3,\x2)
  +{2 } \*\Li(3,1-\x1-\x2)
  +{2 } \*\Li(3,\x1+\x2)
  +{1 \over 2} \*\ln(\x1)\*\Li(2,1-\x1)
  \nn\\&&
  -{1 \over 2} \*\ln(\x1)\*\R(\x1,1-\x1-\x2)
  -{5 \over 2} \*\ln(\x2)\*\R(\x2,1-\x1-\x2)
  \bigg]
  +{1 \over 2} \*{\ln(\x1)\over \x1\*(\x1+\x2)}
  \nn\\&&
  -{1 \over 3}
  \*{(\x1^2+2\*\x1\*\x2+\x2^2-2\*\x1-\x2)\over \x1^2\*(\x1+\x2)^2}\*
  \bigg[
  \ln(1-\x1-\x2)^2
  +\ln(\x1+\x2)\*\ln(1-\x1-\x2)
  \nn\\&&
  +\Li(2,1-\x1-\x2)
  \bigg]
  +{(2-2\*\x2-\x1)\over 12\*\x1^2\*(1-\x2)}\*\R(\x1,\x2)
  +{(4\*\x1+\x2+\x1^2-\x2^2)\over 6\*\x1^2\*(\x1+\x2)^2}
  \*\R(\x1,1-\x1-\x2)
  \nn\\&&
  -{\z2\over 12\*\x1} \*
  \bigg[{1\over (1-\x2)}+{6\over \x1}
  -8\*{(2\*\x1+\x2)\over \x1\*(\x1+\x2)^2}
  \bigg]
  -{\Li(2,1-\x1)\over 12\*\x1\*(\x1+\x2)}\*
  \bigg[5-{(1+\x1)\over (1-\x2)}+{2\over \x1}+{6\over (\x1+\x2)}\bigg]
  \nn\\&&
  -{\Li(2,1-\x2)\over 12\*\x1\*(\x1+\x2)}\*\bigg[
  1-5\*{(1+\x1)\over (1-\x2)}+{10\over \x1}+{6\over (\x1+\x2)}\bigg]
  \nn\\&&
  +{1 \over 18} \*{\R(\x2,1-\x1-\x2)\over \x1\*(\x1+\x2)}\*\bigg[
  17+61\*{\x2\over \x1}-{23\over \x1}-38\*{\x2\*(1-\x2)\over \x1^2}
  +{9\over (\x1+\x2)}\bigg]
  \bigg)
  \nn\\&&
  +{i\*\pi\*\nf \over N}\*\bigg(
  {(\x1^2+2\*\x1\*\x2+\x2^2-2\*\x1-\x2)\over \x1^2\*(\x1+\x2)^2}
  \*\ln(1-\x1-\x2)
  -{1 \over 2} \*{(2-2\*\x2-\x1)\over \x1^2\*(1-\x2)}\*\ln(\x2)
  \nn\\&&
  -{1\over (\x1+\x2)\*\x1}
  -{(1-\x1-\x2)\over\x1^3}\*\R(\x2,1-\x1-\x2)
  \bigg)
  -{1 \over 2} \*i\*\pi\*\nf\*N\*{\ln(\x2) \over \x1\*(1-\x2)}
%%:
%%STOP
.\label{eq:c12nn}
\end{eqnarray}
\def\Li{{\rm Li}}
We have introduced the function $\R(\x1,\x2)$, 
which is well known from ref. \cite{Ellis:1981wv}, 
\begin{equation}
  \label{eq:Rfunction}
\R(\x1,\x2) = \left( {1\over 2} \ln(\x1) \ln(\x2)
                -\ln(\x1) \ln(1-\x1)
                +{1\over 2} \z2-\Li_{2}(\x1) \right) + (\x1\leftrightarrow\x2)\, .
\end{equation}
In addition, it is convenient, to define the symmetric functions
$\RR1(\x1,\x2)$ and $\RR2(\x1,\x2)$ as
\begin{eqnarray}
  \label{eq:RRfunctions}
\RR1(\x1,\x2)  &=& \left(
          \ln(\x1) \Li_{1,1}\left({\x1 \over \x1+\x2},\x1+\x2 \right)
         -{1 \over 2}\ln(1-\x1-\x2) \z2
         -\ln(\x1) \Li_{2}(\x1+\x2)
\right.
\nonumber \\ & &
\left.
         -{1 \over 2}\ln(\x1) \ln(\x2) \ln(1-\x1-\x2)
         -\Li_{1,2}\left({\x1 \over \x1+\x2},\x1+\x2 \right)
\right.
\nonumber \\ & &
\left.
         -\Li_{2,1}\left({\x1 \over \x1+\x2},\x1+\x2 \right)
         +\Li_{3}(\x1+\x2)
       \right) + (\x1\leftrightarrow\x2)\, .
\nonumber \\
\RR2(\x1,\x2)  &=& \left(
          {1 \over 2} \ln(\x1)^2 \ln(1-\x1)
         -{3 \over 4} \ln(\x1)^2 \ln(\x2)
         -{1 \over 4} \ln(\x1) \Li_{2}(1-\x1)
\right.
\nonumber \\ & &
\left.
         -{1 \over 4} \ln(\x2) \Li_{2}(1-\x1)
         -\Li_{3}(1-\x1)
         +\Li_{3}(\x1) \right) + (\x1\leftrightarrow\x2)\, .
\end{eqnarray}

We see in eqs. (\ref{eq:c2nn})--(\ref{eq:c12nn}) 
that the multiple polylogarithms 
all have simple arguments of a particular structure, that 
can easily be continued analytically. 
Details of this procedure are discussed in appendix \ref{sec:analyticcont}.

\section{Conclusions}
\label{sec:conclusions}

Determinations of the strong coupling constant $\alpha_s$ from 
data for $e^+ e^- \rightarrow \mbox{3 jets}$ demand 
next-to-next-to-leading (NNLO) theoretical predictions in 
perturbative QCD.
In this paper we have calculated the fermionic contributions 
to the two-loop amplitude $e^+e^- \to q \qb g$. 
Furthermore, we have obtained the full one-loop amplitude to 
order $\eps^2$ in dimensional regularization, needed for the interference
of the one-loop amplitude with itself.

We have used a systematic, fast and efficient method for the
calculation of loop integrals. 
Our procedure allows for a direct evaluation of all scalar integrals in
arbitrary dimensions and with arbitrary powers of propagators by means 
of nested sums. 
The approach relies on the ability to solve all nested sums 
in terms of multiple polylogarithms, which appear to be the natural 
class of functions for virtual corrections in perturbative QCD.
As a consequence, the results allow for analytic continuation in a
straightforward manner. 
Therefore, they apply also to $(2+1)$-jet production
in deep-inelastic scattering or to the production of a massive vector boson 
in hadron-hadron collisions.

The results presented in this paper represent one contribution to the 
full next-to-next-to-leading order calculation of 
$ e^+ e^- \rightarrow \mbox{3 jets}$. 
At the same time, it provides an important cross check on the results 
recently obtained by Garland et al. \cite{Garland:2002ak} with a completely 
independent method. 
Extending our approach to the calculation of the remaining contributions to the 
two-loop amplitude for $e^+e^- \to q \qb g$, i.e. the terms which are not
enhanced by $n_f$, can be done along the lines of section \ref{sec:method}. 

\subsection*{Acknowledgements}
\label{sec:acknowledgements}
We would like to thank A. Brandenburg for carefully reading the manuscript.

\begin{appendix}
\section{One-loop amplitude}
\label{sec:appendixresults}
Here we list the result for the one-loop function 
$c_{12}^{(1),\fin}$ to order $\epsilon^2$.
The remaining functions $c_{i}^{(1),\fin}$, $i=2,4,6$, are of a similar length, and in order
to save a few trees, we do not list them explicitly here.
All results for the one-loop functions 
$c_{i}^{(1),\fin}$, $i=2,4,6,12$, 
together with the fermionic contributions $c_{i}^{(2),\fin}$, $i=2,4,6,12$
to the two-loop amplitude can be obtained as a FORM file from the preprint server
http://arXiv.org by downloading the source file of this article.
Furthermore they are available upon request from the authors.

\def\Ax2o1mx1{\frac{\x2}{1-\x1}}
\def\A1mx1mx2o1mx1{\frac{1-\x1-\x2}{1-\x1}}
\def\Li(#1,#2){{{\rm Li}_{#1}(#2)}}

\begin{eqnarray}
  &&c_{12}^{(1),\fin}(\x1,\x2) =
%%START
%%c1_{12}=
  N\*(1 + i\*\pi\*\eps) \* {\ln(\x2) \over \x1\*(1-\x2)} 
  +(\eps + i\*\pi\*\eps^2)\*N\* \bigg( 
  {1 \over 2}\*{(6\*\ln(\x2)-\ln(\x2)^2) \over (1-\x2)\*\x1}
  -{\R(\x1,\x2) \over (1-\x1-\x2)\*\x1}  
  \bigg)
  \nn\\&&
  +{(1 + i\*\pi\*\eps) \over N}\* \bigg( 
  {2 \over \x1\*\x2}-{\ln(\x2) \over (1-\x2)\*\x1} - {2 \over (\x1+\x2)\*\x2}
  -2\*{\ln(1-\x1-\x2) \over (\x1+\x2)^2\*\x2} 
  + 2\*{\ln(1-\x1-\x2) \over \x1^2\*\x2}
  \nn\\&&
  -2\*{\ln(1-\x1-\x2)-\ln(\x2) \over \x1^2}
  +2\*{(1-\x1-\x2) \over \x1^3}\*\R(\x2,1-\x1-\x2)
  \bigg)
  \nn\\&&
  +{(\eps + i\*\pi\*\eps^2) \over N}\* \bigg( 
  {(4-\ln(\x1)-\ln(1-\x1-\x2)) \over \x1\*\x2}
  -{(1-\x2) \over \x1^2\*\x2}\*(\ln(1-\x1-\x2)-3)\*\ln(1-\x1-\x2)
  \nn\\&&
  +{1 \over \x1^2}\*(3-\ln(\x2))\*\ln(\x2)
  -{(1-\x1+\x2) \over \x1\*\x2^2}\*\R(\x1,1-\x1-\x2)
  +{(1-3\*\x1-\x2) \over \x1^3} \*\R(\x2,1-\x1-\x2)
  \nn\\&&
  -{1 \over 2}\*{(6\*\ln(\x2)-\ln(\x2)^2) \over (1-\x2)\*\x1}
  -2\*{(2-\ln(1-\x1-\x2)) \over (\x1+\x2)\*\x2}
  -{(4-\ln(1-\x1-\x2)) \over (\x1+\x2)^2\*\x2}\*\ln(1-\x1-\x2)
  \nn\\&&
  +{(1-\x1-\x2) \over \x1^3} \* \bigg[ 
   (3\*\ln(1-\x2)-\ln(1-\x1-\x2))\*\ln(\x2)^2
  -2\*\RR1(1-\x1-\x2,\x2) 
  \nn\\&&
  +(2\*\Li(2,1-\x2)-4\*\z2-\ln(1-\x1-\x2)^2)\*\ln(\x2)
  +6\*\Li(3,1-\x1-\x2)+6\*\Li(3,\x2)  
  \nn\\&&
  +\ln(\x1+\x2)\*\ln(1-\x1-\x2)^2-2\*(\Li(2,1-\x1-\x2)+\z2)\*\ln(1-\x1-\x2)
  \bigg]
  \bigg)
  \nn\\&&
  +\eps^2\*N\* \bigg( 
  {1 \over 6}\*{(-21\*\z2+36-9\*\ln(\x2)+\ln(\x2)^2) \over \x1\*(1-\x2)}\*\ln(\x2) 
  +{1 \over 2}\*{1 \over \x1\*(1-\x1-\x2)}\*\bigg[
  -6\*\Li(3,\x1)
  \nn\\&&
  -6\*\Li(3,\x2)
  -3\*\ln(\x2)^2\*\ln(1-\x2)   
  +(-3\*\ln(1-\x1)+\ln(\x2))\*\ln(\x1)^2+4\*\ln(\x2)\*\z2
  \nn\\&&
  -(2\*\Li(2,1-\x1)-\ln(\x2)^2-4\*\z2)\*\ln(\x1)
  -6\*\R(\x1,\x2)+2\*\RR1(\x1,\x2)-2\*\Li(2,1-\x2)\*\ln(\x2) 
  \bigg] 
  \bigg)
  \nn\\&&
  +{\eps^2 \over N}\* \bigg( 
   {1 \over 6}\*{(21\*\z2-36+9\*\ln(\x2)-\ln(\x2)^2) \over \x1\*(1-\x2)}\*\ln(\x2) 
  -{(1-\x1+3\*\x2) \over \x1\*\x2^2} \*\R(\x1,1-\x1-\x2)
  \nn\\&&
  -{(8-7\*\z2-4\*\ln(1-\x1-\x2)+\ln(1-\x1-\x2)^2) \over (\x1+\x2)\*\x2}
  +{1 \over 3}\*{\ln(1-\x1-\x2) \over (\x1+\x2)^2\*\x2}\*(21\*\z2-24
  \nn\\&&
  +6\*\ln(1-\x1-\x2)-\ln(1-\x1-\x2)^2)
  +{(1-\x1+\x2) \over \x1\*\x2^2} \* \bigg[
  \RR1(\x1,1-\x1-\x2)     
  -\Li(2,1-\x1)\*\ln(\x1)
  \nn\\&&
  +{1 \over 2} \*(\ln(\x1)-\ln(\x1+\x2))\*\ln(1-\x1-\x2)^2
  +2\*\ln(\x1)\*\z2-{3 \over 2} \*\ln(\x1)^2\*\ln(1-\x1)-3\*\Li(3,\x1) 
  \nn\\&&
  +{1 \over 2}\*(2\*\z2+2\*\Li(2,1-\x1-\x2)+\ln(\x1)^2)\*\ln(1-\x1-\x2)
  -3\*\Li(3,1-\x1-\x2)
  \bigg] 
  \nn\\&&
  +{(1-3\*\x1-\x2) \over \x1^3} \* \bigg[ 
  \Li(2,1-\x2)\*\ln(\x2)
  -\RR1(1-\x1-\x2,\x2)
  +3\*\Li(3,1-\x1-\x2)+3\*\Li(3,\x2) 
  \nn\\&&
  +{1 \over 2}\*(\ln(\x1+\x2)-\ln(\x2))\*\ln(1-\x1-\x2)^2
  -\Li(2,1-\x1-\x2)\*\ln(1-\x1-\x2)
  \nn\\&&
  +3\*\R(\x2,1-\x1-\x2)
  -{1 \over 2}\*\ln(\x2)^2\*(\ln(1-\x1-\x2)-3\*\ln(1-\x2))  
  \bigg]
  +{\ln(1-\x1-\x2) \over \x1^2\*\x2}\*\bigg[
  6-7\*\z2
  \nn\\&&
  +{1 \over 3}\*\ln(1-\x1-\x2)^2-{3 \over 2} \*\ln(1-\x1-\x2)\bigg]
  +{1 \over \x1\*\x2}\*\bigg[
  8-7\*\z2-2\*\ln(1-\x1-\x2)+{1 \over 2} \*\ln(\x1)^2
  \nn\\&&
  +{1 \over 2}\*\ln(1-\x1-\x2)^2-2\*\ln(\x1)\bigg]
  +{1 \over \x1^2}\*\bigg[{3 \over  2}\*\ln(1-\x1-\x2)^2+9\*\ln(1-\x1-\x2)\*\z2
  \nn\\&&
  -{1 \over 3}\*\ln(1-\x1-\x2)^3+4\*\R(\x2,1-\x1-\x2)-6\*\ln(1-\x1-\x2)-3\*\ln(\x2)\*\z2+6\*\ln(\x2)
  \nn\\&&
  -{3 \over 2} \*\ln(\x2)^2+{1 \over 3} \*\ln(\x2)^3\bigg]
  +{(1-\x1-\x2) \over \x1^3} \* \bigg[ 
  -2\*\RR1(1-\x1-\x2,\x2)\*\ln(1-\x1-\x2)
  -{3 \over 5}\*\z2^2 
  \nn\\&&
  - (2\*\ln(\x2)+\ln(1-\x1-\x2)
  + 2\*\R(\x1,\x2)+4\*\R(\x1,1-\x1-\x2) 
  + 5\*\R(\x2,1-\x1-\x2))\*\z2
  \nn\\&&
  +12\*\Li(4,1-\x1)
  -14\*\Li(4,1-\x1-\x2)
  -14\*\Li(4,\x2)
  -3\*\Li(2,1-\x1)\*\ln(1-\x1-\x2)^2
  \nn\\&&
  +(2\*\z2-\ln(\x1)\*\ln(1-\x1-\x2)-\Li(2,1-\x1)
  +{1 \over 2}\*\ln(1-\x1-\x2)^2-\Li(2,1-\x2))\*\ln(\x2)^2
  \nn\\&&
    -{1 \over 3}\*\ln(\x1+\x2)\*\ln(1-\x1-\x2)^3
  +\Li(2,1-\x1-\x2)\*\ln(1-\x1-\x2)^2
  +4\*\Lii(1,3,1,1-\x1)
  \nn\\&&
  +{1 \over 3} \*(\ln(1-\x1-\x2)
  - 4\*\ln(1-\x2))\*\ln(\x2)^3
  +\Lii(1,1,\Ax2o1mx1,1-\x1)\*\ln(\x2)^2
  \nn\\&&
  +4\*\Lii(2,2,1,1-\x1)+4\*\Lii(3,1,1,1-\x1)
  +2\*(\Li(3,1-\x1)+\Li(3,1-\x1-\x2))\*\ln(1-\x1-\x2)
  \nn\\&&
  -2\*(\Lii(2,1,1,1-\x1)+\Lii(1,2,1,1-\x1))\*(\ln(1-\x1-\x2)+\ln(\x2))
  +\bigg[
   2\*\Lii(2,1,\Ax2o1mx1,1-\x1)
  \nn\\&&
  -2\*\Li(3,1-\x1) 
  + \ln(1-\x1-\x2)\*\ln(\x1)^2
  -2\*\ln(1-\x1-\x2)\*\Li(2,1-\x1)    
  +2\*\Li(3,\x2)
  \nn\\&&
  +2\*\Lii(1,1,\Ax2o1mx1,1-\x1)\*\ln(1-\x1-\x2)
  +2\*\Lii(1,2,\Ax2o1mx1,1-\x1)
  +{1\over 3}\*\ln(1-\x1-\x2)^3
  \nn\\&&
  -3\*\ln(\x1)\*\ln(1-\x1-\x2)^2
  +2\*\Liii(1,1,1,\Ax2o1mx1,1,1-\x1)
   \bigg]\*\ln(\x2)
  -6\*\Lii(3,1,\Ax2o1mx1,1-\x1)
  \nn\\&&
  -2\*\Liii(2,1,1,\Ax2o1mx1,1,1-\x1)
  -2\*\Liii(1,1,2,\A1mx1mx2o1mx1,1,1-\x1)
  -6\*\Lii(2,2,\Ax2o1mx1,1-\x1)
  \nn\\&&
  -2\*\Liii(1,1,2,\Ax2o1mx1,1,1-\x1)
  -2\*\Liii(1,2,1,\Ax2o1mx1,1,1-\x1)
  -6\*\Lii(2,2,\A1mx1mx2o1mx1,1-\x1)
  \nn\\&&
  -6\*\Lii(1,3,\Ax2o1mx1,1-\x1)
  -2\*\Liii(2,1,1,\A1mx1mx2o1mx1,1,1-\x1)
  -6\*\Lii(1,3,\A1mx1mx2o1mx1,1-\x1)
  \nn\\&&
  -2\*\Liii(1,2,1,\A1mx1mx2o1mx1,1,1-\x1)
  -6\*\Lii(3,1,\A1mx1mx2o1mx1,1-\x1)
  +\bigg[\ln(1-\x1-\x2)^2
  \nn\\&&
  +\ln(\x1)^2-2\*\Li(2,1-\x1-\x2)
  -2\*\ln(\x1+\x2)\*\ln(1-\x1-\x2)
  +6\*\Li(2,1-\x1)\bigg]\*\z2
  \nn\\&&
  +3\*\Lii(1,1,\A1mx1mx2o1mx1,1-\x1)\*\ln(1-\x1-\x2)^2
  +2\*\bigg[
   \Liii(1,1,1,\A1mx1mx2o1mx1,1,1-\x1)
  \nn\\&&
  -\Lii(2,1,\Ax2o1mx1,1-\x1)
  -\Lii(1,2,\Ax2o1mx1,1-\x1)\bigg]\*\ln(1-\x1-\x2)
  \bigg] 
  \bigg)
  %%
  %%
%%:
%%STOP
\, .
\label{eq:c12oneloop}
\end{eqnarray}
\def\Li{{\rm Li}}

\section{Analytic continuation}
\label{sec:analyticcont}
The multiple polylogarithms $\mbox{Li}_{m_k,...,m_1}(y_k,...,y_1)$ are analytic functions 
in $k$ complex variables $y_j$, $j=1,...,k$.
To discuss the branch cuts it is convenient to change the variables according to
\bq
z_j & = & 1 - y_1 y_2 ... y_j,
\eq
and one has
\bq
\mbox{Li}_{m_k,...,m_1}(y_k,...,y_1) & = & 
 \mbox{Li}_{m_k,...,m_1}\left( \frac{1-z_k}{1-z_{k-1}},...,
             \frac{1-z_2}{1-z_{1}},1-z_1 \right).
\eq
The multiple polylogarithms have a representation as nested sums. From 
this represenation one deduces that the multiple polylogarithms are real, 
if all $z_j$ are in the intervall $0 \le z_j \le 2$ and $(m_1,z_1) \neq(1,0)$.
If $z_1$, ..., $z_{j-1}$, $z_{j+1}$, ..., $z_k$ are fixed in this interval,
a given multiple polylogarithm is an analytic function in the remaining variable $z_j$ 
with a branch cut  along the
negative real axis starting from $0$.
We denote by $\mbox{Re}_{z_j}$ and $\mbox{Im}_{z_j}$ the real and imaginary part 
with respect to the variable $z_j$.
In the calculation presented here, the $z_j$'s are ratios of two kinematical invariants:
\bq
z_j & = & \frac{-s_j}{-t_j}
\eq
In fact, the only ratios which occur are:
\bq
x_1 = \frac{-s_{12}}{-s_{123}}, & & 1-x_1 = \frac{-s_{23}-s_{13}}{-s_{123}}, \nonumber \\
x_2 = \frac{-s_{23}}{-s_{123}}, & & 1-x_2 = \frac{-s_{12}-s_{13}}{-s_{123}}, \nonumber \\
1-x_1-x_2 = \frac{-s_{13}}{-s_{123}}, & & x_1+x_2 = \frac{-s_{12}-s_{23}}{-s_{123}}.
\eq
The real and imaginary part of the logarithm is given by
\bq
\mbox{Re}_z \; \mbox{Li}_1(1-z) & = & \ln \left| z \right|, \nonumber \\
\mbox{Im}_z \; \mbox{Li}_1(1-z) & = & - \mbox{Im}\; \ln \left( \frac{-s-i0}{-t-i0} \right)
                               =    \pi \left[ \theta(s) - \theta(t) \right] \label{defim}.
\eq
Eq. (\ref{defim}) defines the sign of the imaginary part for a ratio of two invariants.
All imaginary parts of higher multiple polylogarithms 
can be related to the imaginary part of the logarithm.
They are easily obtained from 
the iterated integral represantation
\bq
\lefteqn{
\mbox{Li}_{m_k,...,m_1}(x_k,...,x_1) \, =} 
\nonumber \\ & & 
 \int\limits_0^{x_1} \left( \frac{dt}{t} \circ \right)^{m_1-1} \frac{dt}{1-t}
 \circ \int\limits_0^{t x_2} \left( \frac{dt}{t} \circ \right)^{m_2-1} \frac{dt}{1-t}
 \circ ...
 \circ \int\limits_0^{t x_k} \left( \frac{dt}{t} \circ \right)^{m_k-1}
 \frac{dt}{1-t}\, , 
\eq
and the fact that
\bq
\mbox{Im}_z  \frac{1}{1-z\pm i0} & = & \mp \pi \frac{\partial}{\partial z} \Theta(z-1).
\eq
The imaginary part in the variable $z_j$ is given by
\bq
\lefteqn{
 \mbox{Im}_{z_j} \; \mbox{Li}_{m_k,...,m_1}\left( \frac{1-z_k}{1-z_{k-1}},...,
             \frac{1-z_2}{1-z_{1}},1-z_1 \right)  
 =  } \nonumber \\
 & & \left[ \mbox{Im} \; \mbox{Li}_1\left(1-z_j\right) \right] \times 
 \int\limits_0^{1-z_1} 
 \left( \frac{dt}{t} \circ \right)^{m_1-1} 
  \frac{dt}{1-t} \circ ...
\nonumber \\
&  & 
 \int\limits_0^{\frac{1-z_j}{1-z_{j-1}} t} 
 \left( \frac{dt}{t} \circ \right)^{m_j-1} 
  \int\limits_0^t dt \left[ \frac{\partial}{\partial t} \theta(t-1) \right]
 \circ ...
 \circ \int\limits_0^{\frac{1-z_k}{1-z_{k-1}} t} \left( \frac{dt}{t} \circ \right)^{m_k-1} \frac{dt}{1-t}.
\eq
Using partial integration, the iterated integral is then related to multiple polylogarithms of reduced
weight.
For the dilogarithm and the trilogarithm one obtains the well-known formulae
\bq
\mbox{Im} \; \mbox{Li}_2(1-z) & = & \ln(1-z) \; \mbox{Im} \; \mbox{Li}_1(1-z), 
\nonumber \\
\mbox{Im} \; \mbox{Li}_3(1-z) & = & \frac{1}{2} \ln^2(1-z) \; \mbox{Im} \; \mbox{Li}_1(1-z).
\eq
In the two-loop amplitude there are also multiple polylogarithms depending on two
variables in the form $\mbox{Li}_{ab}((1-z_2)/(1-z_1),1-z_1)$, where $(a,b) = (1,1)$, $(1,2)$
or $(2,1)$.
In general we have
\bq
{\lefteqn{
\mbox{Li}_{ab}\left( \frac{1-z_2}{1-z_1}, 1-z_1 \right) \,=}} 
\nonumber \\ &  & 
 \mbox{Re}_{z_1} \; \mbox{Re}_{z_2} \; \mbox{Li}_{ab}\left( \frac{1-z_2}{1-z_1}, 1-z_1 \right)
 + i \; \mbox{Re}_{z_1} \; \mbox{Im}_{z_2} \; \mbox{Li}_{ab}\left( \frac{1-z_2}{1-z_1}, 1-z_1 \right)
\nonumber \\ & &
 + i \;\mbox{Im}_{z_1} \; \mbox{Re}_{z_2} \; \mbox{Li}_{ab}\left( \frac{1-z_2}{1-z_1}, 1-z_1 \right)
 - \mbox{Im}_{z_1} \; \mbox{Im}_{z_2} \; \mbox{Li}_{ab}\left( \frac{1-z_2}{1-z_1}, 1-z_1 \right).
\eq
For the imaginary parts we find:
\bq
\lefteqn{
\hspace*{-60mm}
  \mbox{Im}_{z_1} \; \mbox{Li}_{11}\left( \frac{1-z_2}{1-z_1}, 1-z_1 \right)
  =  
    \mbox{Li}_{1}\left( \frac{1-z_2}{1-z_1} \right) \;
    \mbox{Im}_{z_1} \; \mbox{Li}_{1}\left( 1-z_1 \right), 
  } 
\nonumber \\
\lefteqn{
\hspace*{-60mm}
  \mbox{Im}_{z_1} \; \mbox{Li}_{12}\left( \frac{1-z_2}{1-z_1}, 1-z_1 \right)
  =  
    - \mbox{Li}_1(z_1) \mbox{Li}_{1}\left( \frac{1-z_2}{1-z_1} \right) \;
    \mbox{Im}_{z_1} \; \mbox{Li}_{1}\left( 1-z_1 \right), 
  }
\nonumber \\
\lefteqn{
\hspace*{-60mm}
  \mbox{Im}_{z_1} \; \mbox{Li}_{21}\left( \frac{1-z_2}{1-z_1}, 1-z_1 \right)
  =  
    \mbox{Li}_{2}\left( \frac{1-z_2}{1-z_1} \right) 
    \;
    \mbox{Im}_{z_1} \; \mbox{Li}_{1}\left( 1-z_1 \right). 
  }
\eq
\bq
\lefteqn{
 \mbox{Im}_{z_2} \; \mbox{Li}_{11}\left( \frac{1-z_2}{1-z_1}, 1-z_1 \right)
  =  
    \left[ \mbox{Li}_{1}\left( 1-z_1 \right) 
          - \mbox{Li}_{1}\left( \frac{1-z_1}{1-z_2} \right) 
    \right] \;
 \mbox{Im}_{z_2} \; \mbox{Li}_{1}\left( 1-z_2 \right), }
\nonumber \\
\lefteqn{
\mbox{Im}_{z_2} \; \mbox{Li}_{12}\left( \frac{1-z_2}{1-z_1}, 1-z_1 \right) }
  \nonumber \\
  & = & 
       \left[ \mbox{Li}_{2}\left( 1-z_1 \right) 
            - \mbox{Li}_{2}\left( \frac{1-z_1}{1-z_2} \right) 
            + \mbox{Li}_1(z_2) \mbox{Li}_1\left( \frac{1-z_1}{1-z_2} \right)
       \right] \;
 \mbox{Im}_{z_2} \; \mbox{Li}_{1}\left( 1-z_2 \right),
\nonumber \\
\lefteqn{
  \mbox{Im}_{z_2} \; \mbox{Li}_{21}\left( \frac{1-z_2}{1-z_1}, 1-z_1 \right) }
  \nonumber \\
   & = &  
     - \left[ \mbox{Li}_{2}\left( 1-z_1 \right) 
            - \mbox{Li}_{2}\left( \frac{1-z_1}{1-z_2} \right)
            + \mbox{Li}_{1}\left( z_2 \right) \mbox{Li}_{1}\left( 1-z_1 \right)
       \right] \;
 \mbox{Im}_{z_2} \; \mbox{Li}_{1}\left( 1-z_2 \right). 
\eq
\bq
\lefteqn{
\hspace*{-20mm}
  \mbox{Im}_{z_1} \; \mbox{Im}_{z_2} \; \mbox{Li}_{11}\left( \frac{1-z_2}{1-z_1}, 1-z_1 \right)
  =  
    \theta(z_1 - z_2 )
    \; \mbox{Im}_{z_1} \; \mbox{Li}_{1}\left( 1-z_1 \right)
    \; \mbox{Im}_{z_2} \; \mbox{Li}_{1}\left( 1-z_2 \right)
    , 
  } 
\nonumber \\
\lefteqn{
\hspace*{-20mm}
  \mbox{Im}_{z_1} \; \mbox{Im}_{z_2} \; \mbox{Li}_{12}\left( \frac{1-z_2}{1-z_1}, 1-z_1 \right)
  =  
    - \theta(z_1 - z_2 ) \mbox{Li}_1(z_1)
    \; \mbox{Im}_{z_1} \; \mbox{Li}_{1}\left( 1-z_1 \right)
    \; \mbox{Im}_{z_2} \; \mbox{Li}_{1}\left( 1-z_2 \right)
    , 
  } 
\nonumber \\
\lefteqn{
\hspace*{-20mm}
  \mbox{Im}_{z_1} \; \mbox{Im}_{z_2} \; \mbox{Li}_{21}\left( \frac{1-z_2}{1-z_1}, 1-z_1 \right)
  } \nonumber \\
 & = &  
    \theta(z_1 - z_2 ) \left[ \mbox{Li}_1(z_1) - \mbox{Li}_1(z_2) \right]
    \; \mbox{Im}_{z_1} \; \mbox{Li}_{1}\left( 1-z_1 \right)
    \; \mbox{Im}_{z_2} \; \mbox{Li}_{1}\left( 1-z_2 \right)
    . 
\eq
The imaginary parts of the harmonic polylogarithms $\mbox{Li}_{ab}(1,1-z_1)$ can be obtained
from the formulae above by setting $z_2=z_1$. Special care has to be taken for the double imaginary
part. Here one uses
\bq
\theta(x-1) \frac{\partial}{\partial x} \theta(x-1) & = & \frac{1}{2} 
\frac{\partial}{\partial x} \left[ \theta(x-1) \right]^2
\eq
and a factor $1/2$ appears in the final formulae:
\bq
  \lim\limits_{z_2 \rightarrow z_1} 
  \mbox{Im}_{z_1} \; \mbox{Im}_{z_2} \; \mbox{Li}_{11}\left( \frac{1-z_2}{1-z_1}, 1-z_1 \right)
 &  = & 
    \frac{1}{2}
    \left[ \; \mbox{Im}_{z_1} \; \mbox{Li}_{1}\left( 1-z_1 \right) \right]^2
    , 
\nonumber \\
  \lim\limits_{z_2 \rightarrow z_1} 
  \mbox{Im}_{z_1} \; \mbox{Im}_{z_2} \; \mbox{Li}_{12}\left( \frac{1-z_2}{1-z_1}, 1-z_1 \right)
 &  = & 
    - \frac{1}{2} \mbox{Li}_1(z_1)
    \left[ \; \mbox{Im}_{z_1} \; \mbox{Li}_{1}\left( 1-z_1 \right) \right]^2
    , 
\nonumber \\
  \lim\limits_{z_2 \rightarrow z_1} 
  \mbox{Im}_{z_1} \; \mbox{Im}_{z_2} \; \mbox{Li}_{21}\left( \frac{1-z_2}{1-z_1}, 1-z_1 \right)
 & = &  0
    . 
\eq
In the one-loop amplitude we encounter additional multiple polylogarithms.
We discuss here as an example the imaginary parts of 
$\mbox{Li}_{111}\left( (1-z_3)/(1-z_2), (1-z_2)/(1-z_1), 1-z_1 \right)$:
\bq
\lefteqn{
\hspace*{-20mm}
  \mbox{Im}_{z_1} \; \mbox{Li}_{111}\left( \frac{1-z_3}{1-z_2}, \frac{1-z_2}{1-z_1}, 1-z_1 \right)
 =  
 \mbox{Li}_{11}\left( \frac{1-z_3}{1-z_2}, \frac{1-z_2}{1-z_1} \right)
    \; \mbox{Im}_{z_1} \; \mbox{Li}_{1}\left( 1-z_1 \right),
} \nonumber \\
\lefteqn{
\hspace*{-20mm}
  \mbox{Im}_{z_2} \; \mbox{Li}_{111}\left( \frac{1-z_3}{1-z_2}, \frac{1-z_2}{1-z_1}, 1-z_1 \right)
  =  
 \mbox{Li}_{1}\left( \frac{1-z_3}{1-z_2} \right)
    \left[ \mbox{Li}_{1}\left( 1-z_1 \right) 
          - \mbox{Li}_{1}\left( \frac{1-z_1}{1-z_2} \right) 
    \right] 
  } \nonumber \\
  & &  \times 
  \; \mbox{Im}_{z_2} \; \mbox{Li}_{1}\left( 1-z_2 \right),
 \nonumber \\
\lefteqn{
\hspace*{-20mm}
  \mbox{Im}_{z_3} \; \mbox{Li}_{111}\left( \frac{1-z_3}{1-z_2}, \frac{1-z_2}{1-z_1}, 1-z_1 \right)
 =  
 \left\{
 \mbox{Li}_{11}\left( \frac{1-z_2}{1-z_1}, 1-z_1 \right)
 -\mbox{Li}_{11}\left( \frac{1-z_2}{1-z_1}, \frac{1-z_1}{1-z_3} \right)
 \right. } \nonumber \\
& & \left.
 -\mbox{Li}_{1}\left( \frac{1-z_2}{1-z_3} \right)
    \left[ \mbox{Li}_{1}\left( 1-z_1 \right) 
          - \mbox{Li}_{1}\left( \frac{1-z_1}{1-z_3} \right) 
    \right] \right\}
    \; \mbox{Im}_{z_3} \; \mbox{Li}_{1}\left( 1-z_3 \right),
 \nonumber \\
\lefteqn{
\hspace*{-20mm}
  \mbox{Im}_{z_1} \; \mbox{Im}_{z_2} 
  \; \mbox{Li}_{111}\left( \frac{1-z_3}{1-z_2}, \frac{1-z_2}{1-z_1}, 1-z_1 \right)
 =  
 \theta(z_1-z_2)
 \mbox{Li}_{1}\left( \frac{1-z_3}{1-z_2} \right)
    \; \mbox{Im}_{z_1} \; \mbox{Li}_{1}\left( 1-z_1 \right)
} \nonumber \\
& & \times 
    \; \mbox{Im}_{z_2} \; \mbox{Li}_{1}\left( 1-z_2 \right),
\nonumber \\
\lefteqn{
\hspace*{-20mm}
  \mbox{Im}_{z_1} \; \mbox{Im}_{z_3} 
  \; \mbox{Li}_{111}\left( \frac{1-z_3}{1-z_2}, \frac{1-z_2}{1-z_1}, 1-z_1 \right)
 =  
 \theta(z_1-z_3)
 \left[
  \mbox{Li}_{1}\left( \frac{1-z_2}{1-z_1} \right)
 -\mbox{Li}_{1}\left( \frac{1-z_2}{1-z_3} \right)
 \right]
} \nonumber \\
& & \times 
    \; \mbox{Im}_{z_1} \; \mbox{Li}_{1}\left( 1-z_1 \right)
    \; \mbox{Im}_{z_3} \; \mbox{Li}_{1}\left( 1-z_3 \right),
\nonumber \\
\lefteqn{
\hspace*{-20mm}
  \mbox{Im}_{z_2} \; \mbox{Im}_{z_3} 
  \; \mbox{Li}_{111}\left( \frac{1-z_3}{1-z_2}, \frac{1-z_2}{1-z_1}, 1-z_1 \right)
 =  
 \theta(z_2-z_3)
 \left[
  \mbox{Li}_{1}\left( 1-z_1 \right)
 -\mbox{Li}_{1}\left( \frac{1-z_1}{1-z_2} \right)
 \right]
} \nonumber \\
& & \times 
    \; \mbox{Im}_{z_2} \; \mbox{Li}_{1}\left( 1-z_2 \right)
    \; \mbox{Im}_{z_3} \; \mbox{Li}_{1}\left( 1-z_3 \right),
\nonumber \\
\lefteqn{
\hspace*{-20mm}
  \mbox{Im}_{z_1} \; \mbox{Im}_{z_2} \; \mbox{Im}_{z_3} 
  \; \mbox{Li}_{111}\left( \frac{1-z_3}{1-z_2}, \frac{1-z_2}{1-z_1}, 1-z_1 \right)
 =  
 \theta(z_1-z_2) \theta(z_2-z_3)
    \; \mbox{Im}_{z_1} \; \mbox{Li}_{1}\left( 1-z_1 \right)
} \nonumber \\
& & \times 
    \; \mbox{Im}_{z_2} \; \mbox{Li}_{1}\left( 1-z_2 \right)
    \; \mbox{Im}_{z_3} \; \mbox{Li}_{1}\left( 1-z_3 \right).
\eq
From these formulae the imaginary parts of
$\mbox{Li}_{111}\left( (1-z_3)/(1-z_1), 1, 1-z_1 \right)$ can be obtained by
setting $z_2=z_1$. Double imaginary parts in $z_1$ and $z_2$ are given by
\bq
\lefteqn{
\hspace*{-20mm}
  \lim\limits_{z_2 \rightarrow z_1} 
  \mbox{Im}_{z_1} \; \mbox{Im}_{z_2} 
  \; \mbox{Li}_{111}\left( \frac{1-z_3}{1-z_2}, \frac{1-z_2}{1-z_1}, 1-z_1 \right)
 =  
 \frac{1}{2}
 \mbox{Li}_{1}\left( \frac{1-z_3}{1-z_1} \right)
    \left[ \; \mbox{Im}_{z_1} \; \mbox{Li}_{1}\left( 1-z_1 \right) \right]^2,
} \nonumber \\
\lefteqn{
\hspace*{-20mm}
  \lim\limits_{z_2 \rightarrow z_1} 
  \mbox{Im}_{z_1} \; \mbox{Im}_{z_2} \; \mbox{Im}_{z_3} 
  \; \mbox{Li}_{111}\left( \frac{1-z_3}{1-z_2}, \frac{1-z_2}{1-z_1}, 1-z_1 \right)
} \nonumber \\
 & = & 
 \frac{1}{2} \theta(z_1-z_3)
    \left[ \; \mbox{Im}_{z_1} \; \mbox{Li}_{1}\left( 1-z_1 \right) \right]^2
    \; \mbox{Im}_{z_3} \; \mbox{Li}_{1}\left( 1-z_3 \right).
\eq
At weight four there are in addition the multiple polylogarithms with indices
$\mbox{Li}_{22}$, $\mbox{Li}_{13}$, $\mbox{Li}_{31}$, as well as
$\mbox{Li}_{211}$, $\mbox{Li}_{121}$ and $\mbox{Li}_{112}$.
The imaginary parts of those are obtained in complete analogy.

\end{appendix}


\begin{thebibliography}{10}

\bibitem{Dissertori:2001mv}
G.~Dissertori,
\newblock (2001), hep-ex/0105070.
%%CITATION = HEP-EX 0105070;%%

\bibitem{Ellis:1981wv}
R.~K. Ellis, D.~A. Ross, and A.~E. Terrano,
\newblock Nucl. Phys. {\bf B178}, 421 (1981).
%%CITATION = NUPHA,B178,421;%%

\bibitem{Fabricius:1981sx}
K.~Fabricius, I.~Schmitt, G.~Kramer, and G.~Schierholz,
\newblock Zeit. Phys. {\bf C11}, 315 (1981).
%%CITATION = ZEPYA,C11,315;%%

\bibitem{Kunszt:1989km}
Z.~Kunszt, P.~Nason, G.~Marchesini, and B.~R. Webber,
\newblock To appear in the Proceedings of the 1989 LEP Physics Workshop,
  Geneva, Swizterland, Feb 20, 1989.

\bibitem{Giele:1992vf}
W.~T. Giele and E.~W.~N. Glover,
\newblock Phys. Rev. {\bf D46}, 1980 (1992).
%%CITATION = PHRVA,D46,1980;%%

\bibitem{Catani:1997vz}
S.~Catani and M.~H. Seymour,
\newblock Nucl. Phys. {\bf B485}, 291 (1997), hep-ph/9605323.
%%CITATION = HEP-PH 9605323;%%

\bibitem{Rodrigo:1997gy}
G.~Rodrigo, A.~Santamaria, and M.~S. Bilenky,
\newblock Phys. Rev. Lett. {\bf 79}, 193 (1997), hep-ph/9703358.
%%CITATION = HEP-PH 9703358;%%

\bibitem{Rodrigo:1999qg}
G.~Rodrigo, M.~S. Bilenky, and A.~Santamaria,
\newblock Nucl. Phys. {\bf B554}, 257 (1999), hep-ph/9905276.
%%CITATION = HEP-PH 9905276;%%

\bibitem{Bernreuther:1997jn}
W.~Bernreuther, A.~Brandenburg, and P.~Uwer,
\newblock Phys. Rev. Lett. {\bf 79}, 189 (1997), hep-ph/9703305.
%%CITATION = HEP-PH 9703305;%%

\bibitem{Brandenburg:1998pu}
A.~Brandenburg and P.~Uwer,
\newblock Nucl. Phys. {\bf B515}, 279 (1998), hep-ph/9708350.
%%CITATION = HEP-PH 9708350;%%

\bibitem{Nason:1998nw}
P.~Nason and C.~Oleari,
\newblock Nucl. Phys. {\bf B521}, 237 (1998), hep-ph/9709360.
%%CITATION = HEP-PH 9709360;%%

\bibitem{Berends:1989yn}
F.~A. Berends, W.~T. Giele, and H.~Kuijf,
\newblock Nucl. Phys. {\bf B321}, 39 (1989).
%%CITATION = NUPHA,B321,39;%%

\bibitem{Hagiwara:1989pp}
K.~Hagiwara and D.~Zeppenfeld,
\newblock Nucl. Phys. {\bf B313}, 560 (1989).
%%CITATION = NUPHA,B313,560;%%

\bibitem{Bern:1997ka}
Z.~Bern, L.~Dixon, D.~A. Kosower, and S.~Weinzierl,
\newblock Nucl. Phys. {\bf B489}, 3 (1997), hep-ph/9610370.
%%CITATION = NUPHA,B489,3;%%

\bibitem{Bern:1997sc}
Z.~Bern, L.~Dixon, and D.~A. Kosower,
\newblock Nucl. Phys. {\bf B513}, 3 (1998), hep-ph/9708239.
%%CITATION = NUPHA,B513,3;%%

\bibitem{Glover:1997eh}
E.~W.~N. Glover and D.~J. Miller,
\newblock Phys. Lett. {\bf B396}, 257 (1997), hep-ph/9609474.
%%CITATION = PHLTA,B396,257;%%

\bibitem{Campbell:1997tv}
J.~M. Campbell, E.~W.~N. Glover, and D.~J. Miller,
\newblock Phys. Lett. {\bf B409}, 503 (1997), hep-ph/9706297.
%%CITATION = PHLTA,B409,503;%%

\bibitem{Gorishnii:1991vf}
S.~G. Gorishnii, A.~L. Kataev, and S.~A. Larin,
\newblock Phys. Lett. {\bf B259}, 144 (1991).
%%CITATION = PHLTA,B259,144;%%

\bibitem{Surguladze:1991tg}
L.~R. Surguladze and M.~A. Samuel,
\newblock Phys. Rev. Lett. {\bf 66}, 560 (1991).
%%CITATION = PRLTA,66,560;%%

\bibitem{Baikov:2001aa}
P.~A. Baikov, K.~G. Chetyrkin, and J.~H. Kuhn,
\newblock Phys. Rev. Lett. {\bf 88}, 012001 (2002), hep-ph/0108197.
%%CITATION = HEP-PH 0108197;%%

\bibitem{Giele:2002hx}
W.~Giele {\em et~al.},
\newblock (2002), hep-ph/0204316.
%%CITATION = HEP-PH 0204316;%%

\bibitem{Tarasov:1996br}
O.~V. Tarasov,
\newblock Phys. Rev. {\bf D54}, 6479 (1996), hep-th/9606018.
%%CITATION = HEP-TH 9606018;%%

\bibitem{Tarasov:1997kx}
O.~V. Tarasov,
\newblock Nucl. Phys. {\bf B502}, 455 (1997), hep-ph/9703319.
%%CITATION = HEP-PH 9703319;%%

\bibitem{Anastasiou:1999bn}
C.~Anastasiou, E.~W.~N. Glover, and C.~Oleari,
\newblock Nucl. Phys. {\bf B575}, 416 (2000), hep-ph/9912251.
%%CITATION = NUPHA,B575,416;%%

\bibitem{'tHooft:1972fi}
G.~'t~Hooft and M.~Veltman,
\newblock Nucl. Phys. {\bf B44}, 189 (1972).
%%CITATION = NUPHA,B44,189;%%

\bibitem{Chetyrkin:1981qh}
K.~G. Chetyrkin and F.~V. Tkachev,
\newblock Nucl. Phys. {\bf B192}, 159 (1981).
%%CITATION = NUPHA,B192,159;%%

\bibitem{Gehrmann:1999as}
T.~Gehrmann and E.~Remiddi,
\newblock Nucl. Phys. {\bf B580}, 485 (2000), hep-ph/9912329.
%%CITATION = NUPHA,B580,485;%%

\bibitem{Moch:2001zr}
S.~Moch, P.~Uwer, and S.~Weinzierl,
\newblock J.\ Math.\ Phys. {\bf 43}, 3363 (2002), hep-ph/0110083.
%%CITATION = HEP-PH 0110083;%%

\bibitem{Moch:2001bi}
S.~Moch, P.~Uwer, and S.~Weinzierl,
\newblock (2001), hep-ph/0110407.
%%CITATION = HEP-PH 0110407;%%

\bibitem{Goncharov}
A.~B. Goncharov,
\newblock Math. Res. Lett. {\bf 5}, 497 (1998), (available at
  http://www.math.uiuc.edu/K-theory/0297).

\bibitem{Borwein}
J.~M. Borwein, D.~M. Bradley, D.~J. Broadhurst, and P.~Lisonek,
\newblock math.CA/9910045.

\bibitem{Remiddi:1999ew}
E.~Remiddi and J.~A.~M. Vermaseren,
\newblock Int. J. Mod. Phys. {\bf A15}, 725 (2000), hep-ph/9905237.
%%CITATION = IMPAE,A15,725;%%

\bibitem{Garland:2002ak}
L.~W. Garland, T.~Gehrmann, E.~W.~N. Glover, A.~Koukoutsakis, and E.~Remiddi,
\newblock (2002), hep-ph/0206067.
%%CITATION = HEP-PH 0206067;%%

\bibitem{Bollini:1972ui}
C.~G. Bollini and J.~J. Giambiagi,
\newblock Nuovo Cim. {\bf 12B}, 20 (1972).
%%CITATION = NUCIA,12B,20;%%

\bibitem{Collins}
J.~Collins,
\newblock  (Cambridge University Press, 1984).

\bibitem{Kosower:1991ax}
D.~A. Kosower,
\newblock Phys. Lett. {\bf B254}, 439 (1991).
%%CITATION = PHLTA,B254,439;%%

\bibitem{Bern:1992aq}
Z.~Bern and D.~A. Kosower,
\newblock Nucl. Phys. {\bf B379}, 451 (1992).
%%CITATION = NUPHA,B379,451;%%

\bibitem{Weinzierl:1999xb}
S.~Weinzierl,
\newblock (1999), hep-ph/9903380.
%%CITATION = HEP-PH 9903380;%%

\bibitem{Berends:1981rb}
F.~A. Berends, R.~Kleiss, P.~De~Causmaecker, R.~Gastmans, and T.~T. Wu,
\newblock Phys. Lett. {\bf B103}, 124 (1981).
%%CITATION = PHLTA,B103,124;%%

\bibitem{DeCausmaecker:1982bg}
P.~De~Causmaecker, R.~Gastmans, W.~Troost, and T.~T. Wu,
\newblock Nucl. Phys. {\bf B206}, 53 (1982).
%%CITATION = NUPHA,B206,53;%%

\bibitem{Gastmans:1990xh}
R.~Gastmans and T.~T. Wu,
\newblock Oxford, UK: Clarendon (1990) 648 p. (International series of
  monographs on physics, 80).

\bibitem{Gunion:1985vc}
J.~F. Gunion and Z.~Kunszt,
\newblock Phys. Lett. {\bf B161}, 333 (1985).
%%CITATION = PHLTA,B161,333;%%

\bibitem{Kleiss:1985yh}
R.~Kleiss and W.~J. Stirling,
\newblock Nucl. Phys. {\bf B262}, 235 (1985).
%%CITATION = NUPHA,B262,235;%%

\bibitem{Xu:1987xb}
Z.~Xu, D.-H. Zhang, and L.~Chang,
\newblock Nucl. Phys. {\bf B291}, 392 (1987).
%%CITATION = NUPHA,B291,392;%%

\bibitem{Catani:1998bh}
S.~Catani,
\newblock Phys. Lett. {\bf B427}, 161 (1998), hep-ph/9802439.
%%CITATION = PHLTA,B427,161;%%

\bibitem{Garland:2001tf}
L.~W. Garland, T.~Gehrmann, E.~W.~N. Glover, A.~Koukoutsakis, and E.~Remiddi,
\newblock Nucl. Phys. {\bf B627}, 107 (2002), hep-ph/0112081.
%%CITATION = HEP-PH 0112081;%%

\bibitem{Bern:2002tk}
Z.~Bern, A.~De~Freitas, and L.~Dixon,
\newblock JHEP {\bf 03}, 018 (2002), hep-ph/0201161.
%%CITATION = HEP-PH 0201161;%%

\bibitem{Nogueira:1991ex}
P.~Nogueira,
\newblock J. Comput. Phys. {\bf 105}, 279 (1993).
%%CITATION = JCTPA,105,279;%%

\bibitem{Anastasiou:1999cx}
C.~Anastasiou, E.~W.~N. Glover, and C.~Oleari,
\newblock Nucl. Phys. {\bf B565}, 445 (2000), hep-ph/9907523.
%%CITATION = NUPHA,B565,445;%%

\bibitem{Anastasiou:1999ui}
C.~Anastasiou, E.~W.~N. Glover, and C.~Oleari,
\newblock Nucl. Phys. {\bf B572}, 307 (2000), hep-ph/9907494.
%%CITATION = NUPHA,B572,307;%%

\bibitem{Euler}
L.~Euler,
\newblock Novi Comm. Acad. Sci. Petropol. {\bf 20}, 140 (1775).

\bibitem{Zagier}
D.~Zagier,
\newblock First European Congress of Mathematics, Vol. II, Birkhauser, Boston ,
  497 (1994).

\bibitem{Gonzalez-Arroyo:1979df}
A.~Gonzalez-Arroyo, C.~Lopez, and F.~J. Yndurain,
\newblock Nucl. Phys. {\bf B153}, 161 (1979).
%%CITATION = NUPHA,B153,161;%%

\bibitem{Gonzalez-Arroyo:1980he}
A.~Gonzalez-Arroyo and C.~Lopez,
\newblock Nucl. Phys. {\bf B166}, 429 (1980).
%%CITATION = NUPHA,B166,429;%%

\bibitem{Vermaseren:1998uu}
J.~A.~M. Vermaseren,
\newblock Int. J. Mod. Phys. {\bf A14}, 2037 (1999), hep-ph/9806280.
%%CITATION = IMPAE,A14,2037;%%

\bibitem{Blumlein:1998if}
J.~Blumlein and S.~Kurth,
\newblock Phys. Rev. {\bf D60}, 014018 (1999), hep-ph/9810241.
%%CITATION = PHRVA,D60,014018;%%

\bibitem{Kazakov:1988jk}
D.~I. Kazakov and A.~V. Kotikov,
\newblock Nucl. Phys. {\bf B307}, 721 (1988).
%%CITATION = NUPHA,B307,721;%%

\bibitem{Kazakov:1990jm}
D.~I. Kazakov and A.~V. Kotikov,
\newblock Nucl. Phys. {\bf B345}, 299 (1990),
\newblock Erratum.
%%CITATION = NUPHA,B345,299;%%

\bibitem{Moch:1999eb}
S.~Moch and J.~A.~M. Vermaseren,
\newblock Nucl. Phys. {\bf B573}, 853 (2000), hep-ph/9912355.
%%CITATION = NUPHA,B573,853;%%

\bibitem{Vermaseren:2000we}
J.~A.~M. Vermaseren and S.~Moch,
\newblock Nucl. Phys. Proc. Suppl. {\bf 89}, 131 (2000), hep-ph/0004235.
%%CITATION = HEP-PH 0004235;%%

\bibitem{Vermaseren:2000nd}
J.~A.~M. Vermaseren,
\newblock (2000), math-ph/0010025.
%%CITATION = MATH-PH 0010025;%%

\bibitem{Bauer:2000cp}
C.~Bauer, A.~Frink, and R.~Kreckel,
\newblock J. Symbolic Computation {\bf 33}, 1 (2002), cs.sc/0004015.
%%CITATION = CS.SC 0004015;%%

\bibitem{Weinzierl:2002hv}
S.~Weinzierl,
\newblock Comput. Phys. Commun. {\bf 145}, 357 (2002), math-ph/0201011.
%%CITATION = MATH-PH 0201011;%%

\end{thebibliography}
\end{document}